\documentclass[]{IEEEtran}
\usepackage{mathrsfs}
\usepackage{amsfonts}
\usepackage{graphicx,cite,epsfig,amssymb,amsmath}
\usepackage{color,xcolor}
\definecolor{red}{rgb}{1.00, 0.00, 0.00}
\usepackage{pifont}
\usepackage{stmaryrd}
\usepackage{setspace}
 \usepackage{balance} 
\usepackage{cite}
\usepackage{array}
\usepackage{float}
\usepackage{verbatim}
\usepackage{multirow}
\usepackage{booktabs}
\usepackage{makecell}
\usepackage[linesnumbered, ruled]{algorithm2e}
\usepackage{graphicx}
\usepackage{epstopdf}
\usepackage{mathtools}
\usepackage{diagbox}
\usepackage{cases}
\usepackage{amsthm}
\usepackage[colorlinks,linkcolor = black,anchorcolor = black,citecolor = black]{hyperref}

\providecommand{\algorithmname}{Algorithm}
\floatname{algorithm}{\protect\algorithmname}
\newcommand{\bm}[1]{\mbox{\boldmath{$#1$}}}
\newtheorem{thm}{Theorem}

\newtheorem{lem}{Lemma}

\usepackage{arydshln}
\allowdisplaybreaks[4]
\usepackage{wasysym}
\usepackage[caption=false,font=footnotesize]{subfig}

\newcommand{\name}{PrivISAC\xspace}

\ifodd 0
\newcommand{\revhyh}[1]{{\color{blue}#1}}

\else
\newcommand{\revhyh}[1]{#1}

\fi

\begin{document}

\title{Invisible Walls: Privacy-Preserving ISAC Empowered by Reconfigurable Intelligent Surfaces}
\author{\IEEEauthorblockN{Yinghui He, \IEEEmembership{Member,~IEEE}, Long Fan, \IEEEmembership{Member,~IEEE}, Lei Xie, \IEEEmembership{Senior Member,~IEEE},\\and Jun Luo, \IEEEmembership{Fellow,~IEEE}}
\thanks{

Y. He and J. Luo are with the College of Computing and Data Science, Nanyang Technological University, Singapore 639798 (email: yinghui.he@ntu.edu.sg, junluo@ntu.edu.sg).

L. Fan is with the School of Computer Science, Nanjing University of Posts and Telecommunications, Nanjing 210003, China, and also with the State Key Laboratory of Novel Software Technology, Nanjing University, Nanjing 210023, China (email: fanl@njupt.edu.cn)

L. Xie is with the State Key Laboratory for Novel Software Technology, Nanjing University, Nanjing 210023, China (email: lxie@nju.edu.cn).}
}

\maketitle

\begin{abstract}
\revhyh{The environmental and target-related information inherently carried in wireless signals, such as channel state information (CSI), has brought increasing attention to integrated sensing and communication (ISAC). However, it also raises pressing concerns about privacy leakage through eavesdropping.
While existing efforts have attempted to mitigate this issue, they either fail to account for the needs of legitimate communication and sensing or rely on hardware with high cost. To overcome these limitations, we propose \name, a plug-and-play, low-cost solution that leverages reconfigurable intelligent surface (RIS) to protect user privacy while preserving ISAC performance.
At its core, \name constructs a small set of RIS configurations from two beamforming vectors per RIS row and randomly activates one configuration per time slot to perturb CSI and mask sensitive sensing information from eavesdroppers.
The two vectors are designed to maintain similar communication-direction responses while creating distinct sensing-direction responses, thereby preserving transmission quality and masking sensitive CSI from eavesdroppers.
To enable legitimate sensing, we introduce a time-domain masking and demasking method that maps CSI samples to their RIS configurations and removes configuration-induced discrepancies.
We implement \name on commodity wireless devices and conduct extensive experiments. Results show that \name provides \revhyh{effective privacy protection}
while preserving high-quality communication and sensing performance for legitimate receivers.}
\end{abstract}
\begin{IEEEkeywords}
Integrated sensing and communications, privacy protection, reconfigurable intelligent surfaces, beamforming
\end{IEEEkeywords}


\section{Introduction}
Next-generation wireless systems are envisioned to go beyond high-speed data transmission, aiming to enable ubiquitous intelligence and seamless connectivity among all things~\cite{wang2022vision}. A critical step toward this vision is equipping networks with built-in wireless sensing capabilities to perceive their surroundings.
In this context, integrated sensing and communication (ISAC) has been identified by the International Telecommunication Union (ITU) as one of the six key usage scenarios for future wireless networks~\cite{liu2025itu,liu2022integrated,10663521}. Rather than relying on additional physical sensors, ISAC utilizes the wireless signals already transmitted by infrastructure nodes, such as cellular base stations (BSs) and Wi-Fi access points (APs), to perform sensing tasks, particularly by exploiting readily available channel state information (CSI)~\cite{cigno2022integrating,meneghello2023toward,he2024forward}.
As wireless signals interact with surrounding targets and environments through reflection, diffraction, and scattering, the resulting CSI inherently captures information about nearby targets.
By applying advanced signal processing and artificial intelligence (AI) techniques to extract this information, a variety of sensing applications become feasible, including human activity recognition~\cite{he2023integrated}, respiration monitoring~\cite{zeng2020multisense}, localization~\cite{11288062}, and trajectory tracking~\cite{zhang2023csi,zhang2024cross}.

While privacy has always been a critical concern in traditional communication systems~\cite{liu2019safeguarding,lee2023channel,jameel2018comprehensive,kim2021federated,bloch2021overview}, the advent of ISAC introduces new and intensified challenges. Despite its tremendous potential, ISAC inherently embeds sensing capability into wireless signals, which opens up new avenues for privacy breaches. Adversaries can eavesdrop on transmissions and exploit publicly known pilots to extract sensitive, target-related information, resulting in unintended leakage.
For example, the authors in \cite{li2016csi} demonstrate that by sniffing the wireless signals originated from the very device on which the user is typing, an attacker can infer sensitive inputs such as passwords.
Meanwhile, WiKI-Eve~\cite{hu2023password} shows that it is possible to recover private information by intercepting beamforming feedback transmitted by the user device. Importantly, such privacy breaches are not limited to signals emitted by the user's own device. Attackers can also exploit signals from nearby devices or infrastructure to infer sensitive user behavior. For instance, WiKey~\cite{ali2015keystroke} reveals that CSI collected from surrounding devices can be used to infer password inputs, as typing introduces measurable variations in the wireless channel. Similarly, the authors in~\cite{zhu2020tu,xiao2025lend} utilize CSI fluctuations to track a person’s movement trajectory within an enclosed space. Comparable attacks have also been implemented using signals transmitted by cellular BSs~\cite{ling2020spidermon}. 

To address such privacy leakage, prior work has proposed several defenses. One kind of approach leverages the ability of the transmitter. For example, the authors in~\cite{zhu2020tu} propose to vary the transmit power of the source device to introduce artificial fluctuations, but at the cost of degraded communication. 
MIMOCrypt~\cite{luo2024mimocrypt} leverages precoding to offer better trade-offs, but requires multi-antenna setups, making them unsuitable for low-cost, single-antenna Internet of Things (IoT) devices. 
Another line of work introduces external devices to distort sensing. PhyCloak~\cite{qiao2016phycloak} introduces a full-duplex jammer to disrupt sensing links. However, such devices are costly and monopolize the sensing channel, preventing nearby legitimate devices from conducting their own sensing. 
To address this limitation, reconfigurable intelligent surfaces (RIS) have emerged as a promising low-cost alternative~\cite{zheng2019intelligent,liu2022path,do2021multi,9729741}. 
\revhyh{RIS can manipulate the wireless environment without the need for multiple antennas or full-duplex hardware, and its effectiveness has been demonstrated in a variety of communication and sensing applications~\cite{zheng2020intelligent, huang2019reconfigurable, xu2023joint,huang2020reconfigurable,10143420}, including recent beamforming optimization studies in RIS/mmWave ISAC systems~\cite{singh2025geometric,singh2025optimal}, and physical layer security~\cite{10637343,naeem2023security}, such as secure transmission~\cite{niu2024efficient}.} 
By leveraging this characteristic, IRShield~\cite{staat2022irshield} uses RIS to introduce randomized channel variation and confuse attackers. However, it merely randomizes the phase of certain RIS regions, without fully exploiting RIS beamforming to generate more significant perturbations. Moreover, the design overlooks the requirements of legitimate receivers (Rx). 
\revhyh{Thus, as shown in Tab.~\ref{tab:comp_exist}, existing privacy-preserving solutions still fall short of protecting a broad range of deployed devices, especially low-cost IoT devices\footnote{In this work, low-cost IoT devices mainly refer to single-antenna devices with limited hardware, which makes it difficult for them to support device-side privacy defenses.}, while preserving both communication and sensing performance.}

\begin{table}[t]
	\centering
	\caption{\revhyh{Comparison on legitimate sensing/communication support and deployment requirement/overhead. \Circle: unsupported; \LEFTcircle: noticeable loss; \CIRCLE: negligible loss.}}
 \label{tab:comp_exist}
	\vspace{-1ex}
	\setlength{\tabcolsep}{3mm}{
		\begin{tabular}{c >{\centering\arraybackslash}m{1.7cm} >{\centering\arraybackslash}m{0.8cm}  >{\centering\arraybackslash}m{0.8cm} >{\centering\arraybackslash}m{1.3cm}}
			\toprule
			  & Core  &  Sen. &  Comm. & Req./Cost \\
		      \hline
              \hline
		  [21]  & Power & \LEFTcircle & \LEFTcircle  & No \\
          \hline
          \!\![24]\!\!  & Precoding   & \CIRCLE & \CIRCLE & Multiple antennas\\
          \hline
          \!\![25]\!\!  &  FD jammer &\CIRCLE  & \LEFTcircle  & High\\
          \hline
          \!\![38]\!\!  & RIS  & \Circle & \CIRCLE  & Low \\
          \hline
          \!\!Our\!\!  & RIS & \CIRCLE & \CIRCLE & Low  \\
			\bottomrule
	\end{tabular}}
    \vspace{-2ex}
\end{table}





\begin{figure}[t]
    \centering
    \setlength{\abovecaptionskip}{6pt}
    \includegraphics[width=0.98\linewidth]{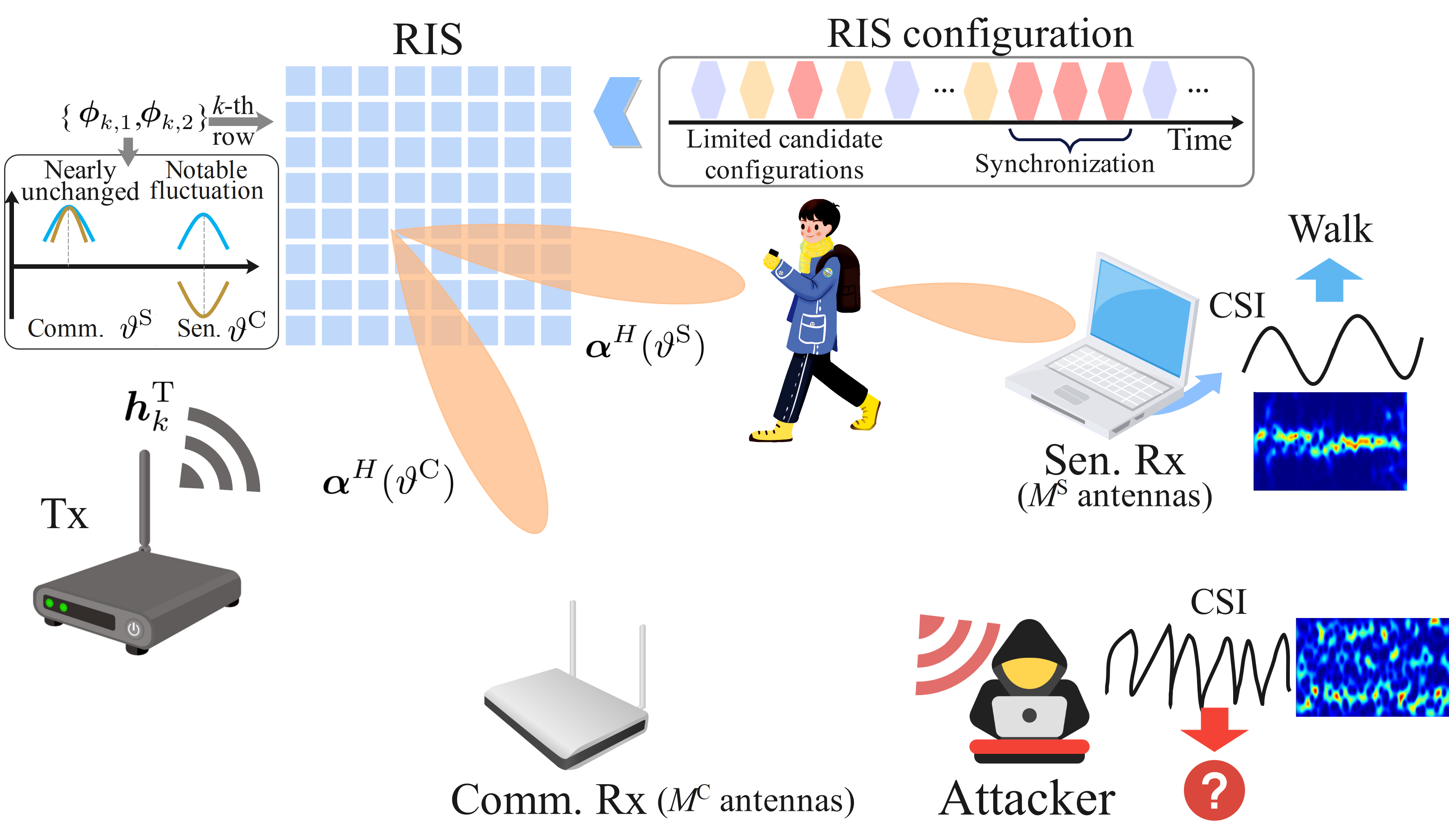}
    \vspace{-0.5em}
    \caption{\name: RIS is leveraged to achieve high-performance ISAC while preserving privacy.}
    \label{fig:intro}
    \vspace{-3ex}
\end{figure}

To bridge this gap, we aim to further exploit the beamforming capability of RIS, rather than only perturbing a small portion of the RIS as in~\cite{staat2022irshield}. To this end, we propose \name, a privacy-aware ISAC system empowered by RIS, as illustrated in Fig.~\ref{fig:intro}.
Specifically, for each row of the RIS, we configure a pair of beamforming vectors. These two vectors are designed to generate significantly different signals in the sensing direction, while producing nearly identical gains in the communication direction. Consequently, when switching between the two vectors for any given row, the sensing signal exhibits notable fluctuations, whereas the communication performance remains nearly unchanged. However, conventional beamforming approaches are not able to achieve such a property. To address this challenge, we formulate a joint optimization problem that jointly designs beamforming vector pairs for all RIS rows.
The objective function considers both privacy-preserving perturbations and communication performance. 
By solving the problem, we propose a beamforming design algorithm under the block coordinate descent (BCD) framework~\cite{bertsekas1999nonlinear}. By partitioning the optimization variables into multiple blocks and optimally updating each block, the algorithm guarantees convergence to a stable solution.
Moreover, since practical RIS implementations typically adopt 1-bit phase quantization, we further extend the proposed algorithm by incorporating a relaxation-and-penalty approach. This design ensures that the optimization procedure remains tractable and converges stably, even under the strict 1-bit constraint. 

If the beamforming vectors for each RIS row were chosen in a completely random manner, then although an illegitimate eavesdropper would be prevented from performing reliable sensing, the legitimate sensing Rx would also struggle to extract meaningful information. This limitation arises primarily because the RIS is a passive device, and thus it cannot actively cancel its own perturbations in the way that the full-duplex jammer can~\cite{qiao2016phycloak}. In fact, the excessive randomness in the RIS configuration space is unnecessary, since our carefully designed beamforming vectors already guarantee sufficient discrepancy in the sensing direction. Motivated by this, we propose a time-domain masking and demasking method. Instead of drawing from the full set of possible configurations, we randomly select a small subset of candidate configurations and then, at each time slot, randomly activate one of them.
This restriction to a limited set enables the legitimate Rx to exploit channel coherence time to reliably estimate configuration-induced effects. Specifically, to ensure that legitimate Rx can correctly associate the measured CSI with the RIS configuration, we embed several consecutive fixed configurations into the sequence. By identifying these fixed patterns, legitimate Rx can achieve precise synchronization with the RIS and further map CSI samples to their corresponding configurations. By leveraging the fact that CSI remains nearly constant within the channel coherence time, the Rx can estimate the relative gain variations introduced by different configurations and compensate for them, 
thereby restoring the CSI sequence and ensuring robust sensing for legitimate Rx.

In summary, we make the following major contributions:
\begin{itemize}
    \item To address the lack of privacy protection in ISAC, we present \name, a RIS-enabled system that ensures high sensing and communication performance while preserving privacy, even on low-cost IoT devices.
    \item We propose an RIS beamforming design algorithm that maintains stable, high-throughput communication while introducing significant fluctuations in the sensing direction to obfuscate sensing information.
    \item We develop a time-domain masking and demasking method that preserves privacy, while enabling legitimate sensing Rx to reliably extract target information.
    \item We implement \name on commodity devices, and extensive experiments confirm its ability to deliver high ISAC performance alongside \revhyh{effective privacy protection}. 
\end{itemize}

The rest of the paper is organized as follows. Section~\ref{sec:background} introduces the attacker model, system model, and presents a feasibility study. Section~\ref{sec:beamforming} formulates the optimization problem and presents a BCD-based beamforming design algorithm for RIS. Section~\ref{sec:design} describes the workflow of \name, including the proposed time-domain masking and demasking method.
Sections~\ref{sec:experiment} and~\ref{sec:result} detail the experiment setup and results. 
Section~\ref{sec:conclusion} concludes the paper.

\textit{Notations:} Scalars are denoted by lower case, vectors are denoted by boldface lower case, and matrices are denoted by boldface upper case. $(\cdot)^*$, $(\cdot)^T$, and $(\cdot)^H$ denote complex conjugate, transpose, and Hermitian transpose, respectively. 
For a vector $\bm{a}$, $\text{Diag}(\bm{a})$ denotes a diagonal matrix with each diagonal element being the corresponding element in $\bm{a}$, $||\bm{a}||$ represents its Euclidean norm, and $\bm{a}[n]$ represents the $n$-th element in $\bm{a}$.
$|\cdot|$ represents the absolute value of a complex scalar. 
$\mathcal{R}\left\{\cdot\right\}$ denotes the real value of a complex scalar.
$\mathbb{C}^{m\times n}$ ($\mathbb{R}^{m\times n}$) denotes the space of $m\times n$ complex (real) matrix. $\mathbb{Z}$ denotes a set of integers.


\section{Preliminary and Motivation} \label{sec:background}

In this section, we first introduce the attack model and the system model with RIS, and then present a feasibility study to motivate the design of \name.  

\subsection{Threat Model} \label{sec:threat_model}

\revhyh{In this paper, as shown in Fig.~\ref{fig:intro}, we consider a constrained and challenging privacy-protection setting, where a single-antenna transmitter (Tx) continuously sends data packets to a communication Rx with $M^{\mathrm{C}}$ antennas to maintain a data link.}
Simultaneously, the transmitted packets are also used for sensing: a separate sensing Rx captures the packets and measures the CSI between the Tx and itself to enable continuous human activity sensing within the environment. Due to hardware constraints in commercial devices, the number of antennas on the sensing Rx, denoted by $M^{\mathrm{S}}$, is typically no more than three. Moreover, we assume that the Tx has knowledge of the CSI between itself and the communication Rx, which can be obtained through standard channel feedback mechanisms~\cite{love2008overview}. In addition, the Tx is also aware of the relative spatial positions of the sensing target, which can be estimated from CSI using existing localization algorithms.

Since the pilot signals transmitted by the Tx are publicly known, an attacker can sniff these signals and obtain the corresponding CSI, even without being registered or authenticated as a legitimate Rx. Specifically, we consider an attacker located within the scenario and it passively captures the transmitted packets using commodity wireless devices or software-defined radios (SDRs) to extract CSI that contains information related to the target.
By analyzing the acquired CSI, the attacker can infer the target's behavior and obtain sensitive information such as typed passwords. In this work, we assume that the attacker possesses the following capabilities~\cite{luo2024mimocrypt}:\\
\noindent $\bullet$ \textbf{Location flexibility}. We assume that the attacker can pre-deploy a sniffing device at any location within the scenario, including positions that coincide with either the communication or sensing Rx. The attacker is free to choose an optimal placement to maximize the success probability of the attack. \\
\noindent $\bullet$ \revhyh{\textbf{Multiple antennas}. We consider that the attacker, relying on commodity wireless devices or SDR platforms, typically has no more than three antennas. Nonetheless, in our experiments, we further evaluate a stronger attacker model where multiple wireless devices are aggregated to form a larger antenna array, and verify that the attacker still fails to extract sensing information, even under this enhanced setup.}\\
\noindent $\bullet$ \textbf{Model knowledge}. We assume that the attacker has access to the same pre-trained sensing model as the legitimate sensing Rx. Upon collecting CSI, the attacker can directly apply this model to obtain privacy. \revhyh{Moreover, we assume that the masking and demasking algorithms are public to the attacker, while the secret key is only shared by legitimate devices.} This assumption ensures that privacy protection achieved is attributed to our design, rather than relying on limitations at the model level.


\subsection{System Model and Goals} \label{sec:system_goal}

An RIS is a two-dimensional programmable structure composed of numerous small and controllable reflecting elements. By adjusting the voltage applied to each element, the phase of the reflected wireless signal can be modified, enabling dynamic control and configuration of the wireless channel. As a result, when an RIS is integrated into a wireless system, the CSI observed at the Rx is influenced by the RIS. In this work, we exploit this property of RIS to enhance privacy protection. Specifically, we deploy the RIS near the Tx and use a directional antenna to steer the Tx's signal toward the RIS, as shown in Fig.~\ref{fig:intro}.
When the RIS has a size of $K \times N$ elements, the diagonal passive beamforming matrix of the $k$-th row of the RIS is denoted by $\bm{\Phi}_k=\mathrm{Diag}([e^{j\psi_{k,1}}, \cdots, e^{j\psi_{k,N}}])\in\mathbb{C}^{N\times N}$ with $\psi_{k,n}$ being the phase of the $(k,n)$-th reflecting element. 

Let $\bm{h}^{\mathrm{T}}_{k} \in \mathbb{C}^{N \times 1}$ denote the wireless channel between the Tx and the $k$-th row of the RIS.
For the communication link, the channel from the $k$-th row of the RIS to the communication Rx is denoted by $\bm{G}^{\mathrm{C}}_{k} \in \mathbb{C}^{M^{\mathrm{C}} \times N}$, and then the channel between the Tx and the communication Rx can be expressed as 
\begin{equation}
    \bm{h}^{\mathrm{Com}} = \sum_{k=1}^{K}  (\bm{G}^{\mathrm{C}}_{k} \bm{\Phi}_k\bm{h}^{\mathrm{T}}_{k} ). \label{eq:channel_com}
\end{equation}
Given the presence of a strong line-of-sight (LoS) path between the RIS and the communication Rx\footnote{We only require LoS on the Tx-RIS and RIS-target/communication-Rx links to ensure that the RIS can effectively shape the propagation. 
}, the channel $\bm{G}^{\mathrm{C}}_{k}$ is primarily dominated by this LoS component with $a_k^{\mathrm{C}}$ being the path loss, $\theta^{\mathrm{C}}$ being the angle of arrival (AoA), and $\vartheta^{\mathrm{C}}$ being the angle of departure (AoD). Then, it can be reasonably approximated as: 
$\bm{G}^{\mathrm{C}}_{k} = a_k^{\mathrm{C}} \bm{\alpha}(\theta^{\mathrm{C}}) \bm{\alpha}^H(\vartheta^{\mathrm{C}})$,
where $\bm{\alpha}(\cdot)$ is the steering vector. Meanwhile, by defining beamforming vector $\bm{\phi}_k \triangleq  [e^{j\psi_{k,1}}, \cdots, e^{j\psi_{k,N}}]^T\in\mathbb{C}^{N\times 1}$ for $k$-th row of the RIS and $\bm{h}^{\mathrm{C}}_{k}  \triangleq (  a_k^{\mathrm{C}} \bm{\alpha}^H(\vartheta^{\mathrm{C}}) \mathrm{Diag}\{\bm{h}^{\mathrm{T}}_{k}\})^H \in \mathbb{C}^{ N \times 1}$, the channel can be rewritten as 
\begin{equation}
    \bm{h}^{\mathrm{Com}}   = \bm{\alpha}(\theta^{\mathrm{C}}) \left(\sum_{k=1}^{K}   (\bm{h}^{\mathrm{C}}_{k})^H \bm{\phi}_k\right). 
\end{equation}
Since $\bm{\alpha}^H(\theta^{\mathrm{C}}) \bm{\alpha}(\theta^{\mathrm{C}}) = M^{\mathrm{C}}$, the communication signal-to-noise ratio (SNR) can be derived as 
\begin{equation}
    \mathrm{SNR}^{\mathrm{Com}} = \frac{ M^{\mathrm{C}} ||\sum_{k=1}^{K}  (\bm{h}^{\mathrm{C}}_{k} )^H\bm{\phi}_k||^2P^{\mathrm{T}}}{\sigma^2},
\end{equation}
where $P^{\mathrm{T}}$ denotes the transmit power at the Tx, and $\sigma^2$ is the power of the complex Gaussian noise at the Rx.


\revhyh{Similarly, for the sensing link, the channel from the $k$-th row of the RIS to the sensing Rx is denoted by $\bm{G}^{\mathrm{S}}_{k} \in \mathbb{C}^{M^{\mathrm{S}} \times N}$.} It mainly contains two parts: the dynamic path $\bm{G}^{\mathrm{S,S}}_k$ related to the sensing target and the static part $\bm{G}^{\mathrm{S,O}}_k$ consisting of other paths, i.e., $\bm{G}^{\mathrm{S}}_{k}=\bm{G}^{\mathrm{S,S}}_k +\bm{G}^{\mathrm{S,O}}_k$. Moreover, the former can be expressed as 
$\bm{G}^{\mathrm{S,S}}_k = a_k^{\mathrm{S}} \bm{\alpha}(\theta^{\mathrm{S}}) \bm{\alpha}^H(\vartheta^{\mathrm{S}})$, 
where $a_k^{\mathrm{S}}$ is the path loss, $\theta^{\mathrm{S}}$ is the AoA, and $\vartheta^{\mathrm{S}}$ is the AoD.
Thus, the CSI measured at the sensing Rx is 
\begin{align} 
    \bm{h}^{\mathrm{Sen}} =& \sum_{k=1}^K (  (\bm{G}^{\mathrm{S,S}}_k +\bm{G}^{\mathrm{S,O}}_k) \bm{\Phi}_k\bm{h}^{\mathrm{T}}_{k}). \label{eq:channel_sen}
\end{align}
By defining $\bm{h}^{\mathrm{S}}_{k}  \triangleq (\bm{\alpha}(\vartheta^{\mathrm{S}})^H \mathrm{Diag}\{\bm{h}^{\mathrm{T}}_{k}\} )^H$, the transmit power towards the direction of the sensing target (i.e., $\vartheta^{\mathrm{S}}$)  for the $k$-th RIS row can be derived as 
\begin{eqnarray} \label{eq:sen_power}
    P^{\mathrm{Sen}}_k = ||  (\bm{h}^{\mathrm{S}}_{k})^H  \bm{\phi}_k ||^2 P^{\mathrm{T}}.
\end{eqnarray}


From equations~\eqref{eq:channel_sen} and~\eqref{eq:sen_power}, it is evident that continuously varying $\bm{\phi}_{k}$ (i.e., $\bm{\Phi}_{k}$) can introduce additional fluctuations in the CSI and received power at the sensing Rx. 
Based on this insight, we can configure each row of the RIS with two distinct passive beamforming vectors\footnote{The reason for selecting two vectors is that they can be designed to produce signals with similar amplitude but opposite phases in the sensing direction, thereby maximizing introduced perturbation.}, denoted by $\bm{\phi}_{k,1}$ and $\bm{\phi}_{k,2}$, and 
\revhyh{further generate different RIS configurations by randomly selecting one beamforming vector for each RIS row, with $N^{\mathrm{R}}$ being the number of configurations.}
Switching between those configurations in a randomized manner can intentionally induce additional randomness into the wireless channel.
Specifically, we aim to achieve the following goals:
\begin{itemize}
    \item Ensure that the Tx maintains a stable and high-speed link to the communication Rx. 
    \item \revhyh{Reduce the leakage of the target through CSI, regardless of the attacker's location.}
    \item Enable the legitimate sensing Rx to extract target-related information from the CSI using a shared key to realize a high sensing performance. 
\end{itemize}

\subsection{Feasibility Study and Motivation} \label{sec:feasibility}

\begin{figure}[t]
    \centering
    \subfloat[Amplitude]{
    \centering
    \includegraphics[width=0.49\linewidth]{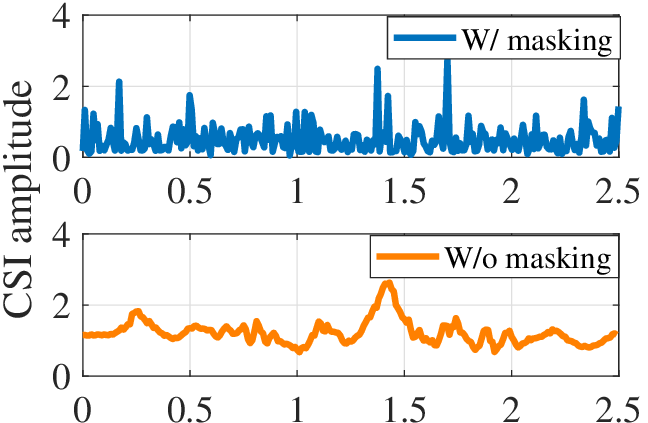}
    \label{fig:csi_a}
    }
    \subfloat[Phase]{
    \centering
    \includegraphics[width=0.49\linewidth]{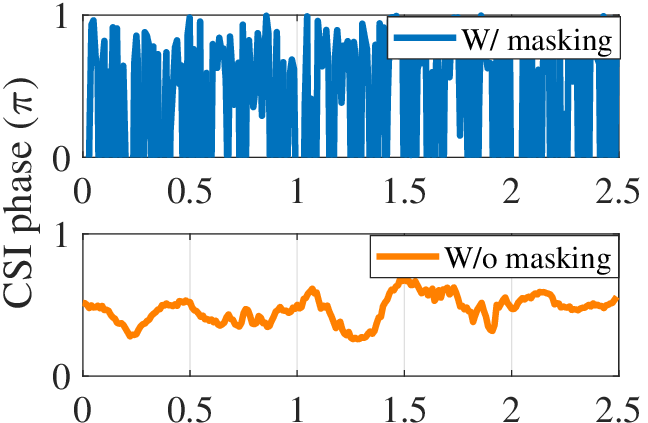}
    \label{fig:csi_p}
    }\\
    \vspace{-2ex}
    \subfloat[W/ masking]{
    \centering
    \includegraphics[width=0.49\linewidth]{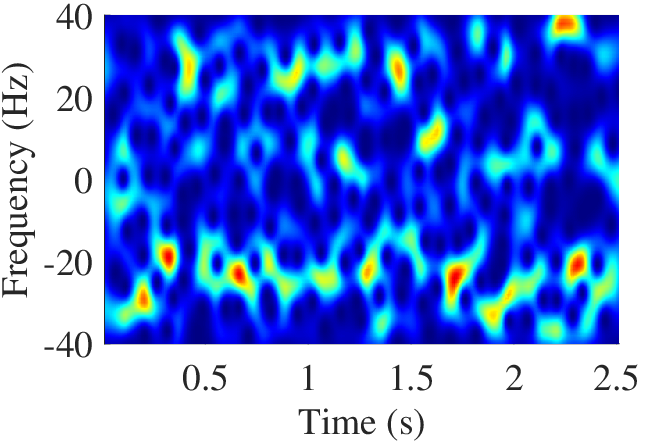}
    \label{fig:time_fre_w}
    }
    \subfloat[W/o masking]{
    \centering
    \includegraphics[width=0.49\linewidth]{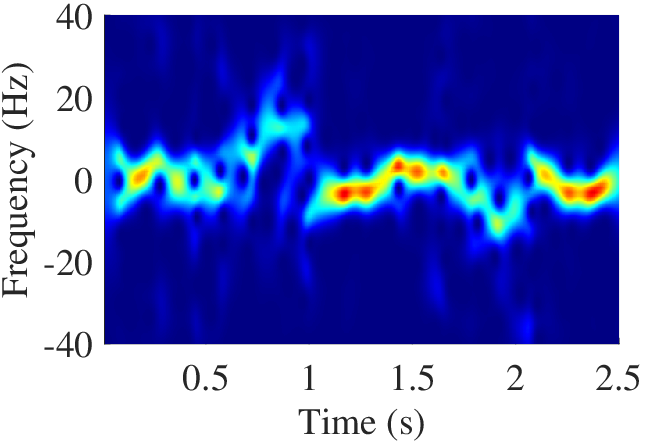}
    \label{fig:time_fre_wo}
    }
    \vspace{-1ex}
    \caption{(a) Amplitude, (b) phase, and (c)-(d) time-frequency analysis of the CSI with and without masking using RIS.}
    \label{fig:csi_pre}
    \vspace{-1.5em}
\end{figure}

We hereby leverage simple experiments to further motivate the design of \name. 
Specifically, we conduct a proof-of-concept experiment using Intel 5300 WiFi network interface cards (NICs). The Tx is equipped with a single antenna to continuously send data packets, while an $8 \times 16$ RIS is deployed to introduce additional variations. A sensing Rx equipped with another Intel 5300 NIC collects the CSI, and the third NIC is deployed as the communication Rx.
In the experiment, a human target repeatedly performs ``slide'' gesture. Note that the detailed experiment setup and layout can be found in Section~\ref{sec:experiment}.
To maximize artificial CSI variation via RIS phase manipulation, we first generate the beamforming vector $\bm{\phi}_{k,1}$ by maximizing the received sensing power $P^{\text{Sen}}_k$. We then construct the second vector $\bm{\phi}_{k,2}$ by adding an additional phase shift of $\pi$ to each element of $\bm{\phi}_{k,1}$, i.e., $\bm{\phi}_{k,2} = -\bm{\phi}_{k,1}$. This design ensures that the two vectors yield the maximum difference in the sensing direction. To introduce artificial fluctuations, we generate four different RIS configurations by randomly selecting one beamforming vector for each RIS row, and further randomly activate one of them at every time slot.

\subsubsection{Privacy Protection}
The measured CSI at the sensing Rx is plotted in Fig.~\ref{fig:csi_pre}. It can be clearly observed that, in the case without RIS, both the amplitude and phase of the CSI exhibit specific patterns, indicating the presence of gesture-related information from the CSI. In contrast, when RIS is applied, the CSI amplitude and phase appear random and disordered, effectively masking the sensing information and preventing potential eavesdropping by attackers.
To further validate the effectiveness of our approach, we also present time-frequency analysis results. As shown in Figs.~\ref{fig:time_fre_w} and~\ref{fig:time_fre_wo}, after masking, the original frequency components are significantly disrupted, making it difficult for an attacker to extract meaningful information. \textit{The above results demonstrate that, when the two beamforming vectors of each RIS row are designed to yield sufficiently pronounced distinctions in the sensing direction, a limited set of RIS configurations (e.g., number being four) is necessary. Randomly switching among them is sufficient to introduce substantial perturbations for privacy protection.}


\subsubsection{Communication Performance} \label{sssec:comm}
\begin{figure}[t]
    \centering
    \setlength{\abovecaptionskip}{6pt}
    \subfloat[Successful ratio]{
        \centering
        \includegraphics[width=0.49\linewidth]{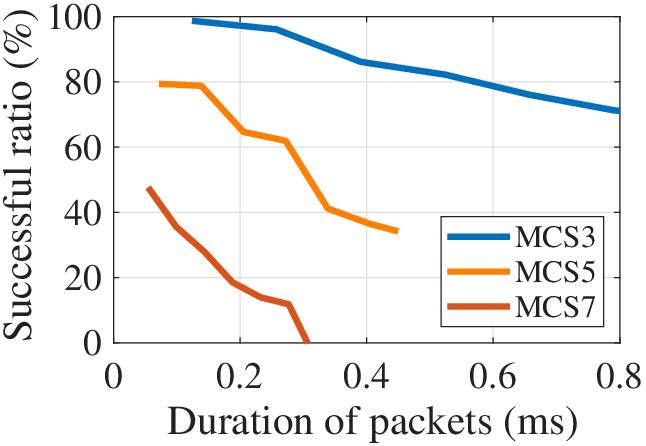}
        \label{fig:comm_rate}
    }
    \subfloat[RSSI]{
        \centering
        \includegraphics[width=0.49\linewidth]{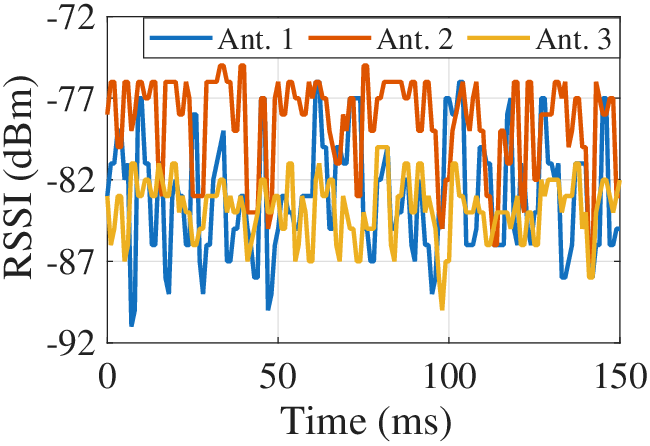}
        \label{fig:comm_rssi}
    }
    \vspace{-1ex}
    \caption{The communication performance: (a) successful transmission ratio under different MCS indices and (b) received signal strength indicator (RSSI) of three antennas.}
    \label{fig:comm_res}
   \vspace{-2ex}
    \centering
    \setlength{\abovecaptionskip}{6pt}
    \subfloat[{Amplitude and phase}]{
        \centering
        \includegraphics[width=0.45\linewidth]{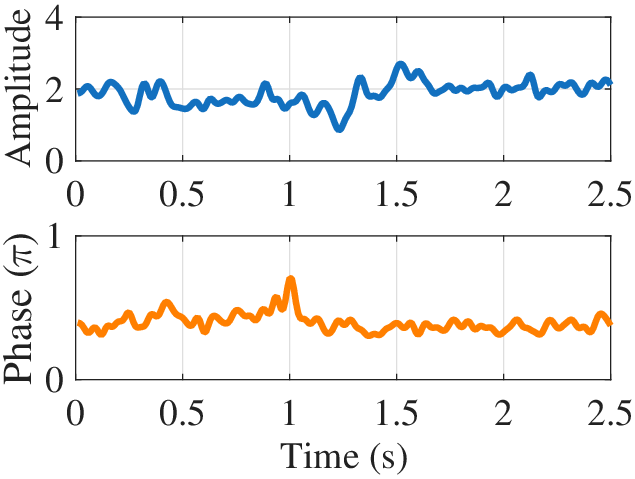}
        \label{fig:time_fre_com}
    }
    \subfloat[Time-frequency analysis]{
        \centering
        \includegraphics[width=0.49\linewidth]{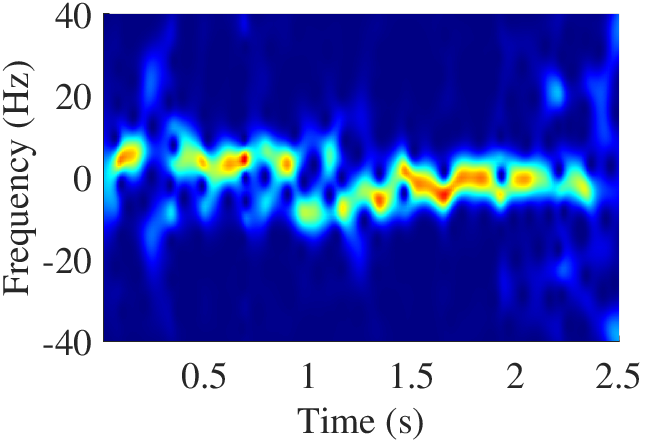}
        \label{fig:time_fre_de}
    }
     \vspace{-1ex}
    \caption{{The CSI after reconstruction: (a) amplitude and phase and (b) time-frequency analysis.}}
    \label{fig:time_fre}
    \vspace{-1.5em}
\end{figure}

Fig.~\ref{fig:comm_rate} shows the ratio of successful packet transmission (i.e., packets correctly decoded by the Rx) under different modulation and coding scheme (MCS) indices as a function of packet duration.
As the figure illustrates, the successful transmission ratio decreases with increasing packet duration. This is because longer packets are more likely to experience RIS configuration switching during transmission, leading to a mismatch between the CSI estimated from the preamble and the actual channel during data decoding, ultimately causing packet failures. \textit{This underscores the need for a robust RIS switching strategy.}
Furthermore, as shown in Fig.~\ref{fig:comm_rssi}, the received signal strength at the communication Rx fluctuates significantly over time, indicating substantial instability during transmission. 
This instability stems from the fact that the beamforming vector design does not account for communication requirements. As a result, randomly switching between such vectors inevitably introduces significant SNR fluctuations at the communication Rx, and may even cause link interruptions.
\textit{Therefore, a novel RIS beamforming design is required, one that ensures a stable communication channel while simultaneously introducing sufficient discrepancy in the sensing direction.}

\subsubsection{Sensing of Legitimate Rx} \label{sssec:sen_rx}
To ensure that the legitimate Rx can perform sensing accurately, it is essential to mitigate the interference introduced by RIS configuration switching. In theory, an optimal solution would be to estimate the channel between each RIS row and the Rx individually. However, this is impractical since RIS is a passive device and cannot provide independent channel measurements, unlike the full-duplex transceivers used in~\cite{qiao2016phycloak}. 
In fact, since the CSI corresponds to a limited set of RIS configurations, once the received CSI samples can be correctly associated with their respective configurations, and the discrepancies among configurations (introduced by RIS beamforming gains) are compensated, an effective CSI sequence can be reconstructed for reliable sensing.
The CSI reconstructed with this approach is illustrated in Fig.~\ref{fig:time_fre}. Compared with the CSI in Fig.~\ref{fig:csi_pre}, the reconstructed CSI still retains identifiable patterns that are usable for sensing.
\textit{However, to make this approach effective, two key issues must be addressed: (1) the Rx must be synchronized with the RIS to ensure that the acquired CSI can be correctly aligned with the underlying configurations, and (2) the discrepancies introduced by different configurations must be compensated to eliminate their impact on sensing.} To this end, we will propose a dedicated time-domain masking and demasking method.


In the following, we first present a novel beamforming design for RIS in Section~\ref{sec:beamforming}, followed by the system design of \name in Section~\ref{sec:design}, including an RIS switching strategy and a time-domain masking and demasking method.

\section{Privacy-Preserving RIS Beamforming Design} \label{sec:beamforming}

In this section, we formulate an optimization problem for RIS beamforming design and propose a BCD-based algorithm.

\subsection{Problem Formulation}

\begin{figure}[t]
    \centering
    \includegraphics[width=0.92\linewidth]{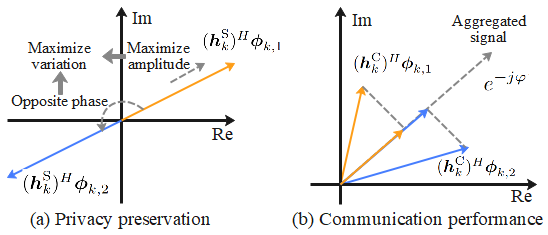}
    \vspace{-2ex}
    \caption{Two objectives for problem formulation.}
    \label{fig:opt}
    \vspace{-1.4em}
\end{figure}

As demonstrated in Section~\ref{sec:feasibility}, a straightforward beamforming vector that maximizes artificial CSI variation is unfriendly to communication performance. Therefore, a more sophisticated beamforming design is required. Specifically, for each RIS row, two beamforming vectors, $\bm{\phi}_{k,1}$ and $\bm{\phi}_{k,2}$, should be designed to achieve the following two objectives:

\noindent $\bullet$ \textbf{Privacy preservation:} As shown in Fig.~\ref{fig:opt}(a), to induce significant CSI fluctuations, both vectors should concentrate signal power toward the sensing direction. \revhyh{This can be achieved by maximizing the aggregate sensing gain: $\max || (\bm{h}^\mathrm{S}_k)^H \bm{\phi}_{k,1} ||^2 + || (\bm{h}^\mathrm{S}_k)^H \bm{\phi}_{k,2} ||^2$. Note that a larger sensing gain also benefits the legitimate sensing user by increasing the received sensing SNR. Meanwhile, the signals produced by the two vectors in the sensing direction should have comparable amplitudes and exhibit near-opposite phases for further maximizing the variation. This can be approximated by minimizing $ || (\bm{h}^\mathrm{S}_k)^H \bm{\phi}_{k,1} + (\bm{h}^\mathrm{S}_k)^H \bm{\phi}_{k,2} ||^2$.}
    
\noindent  $\bullet$ \textbf{Communication performance:}
\revhyh{we aim to encourage both RIS vectors to maintain strong in-phase responses in the communication direction, so that the communication beam remains as stable and high-gain as possible throughout the switching process.}
Since each row randomly selects one of the two vectors, we aim to optimize the worst-case communication performance over all possible combinations:
\begin{equation} \label{eq:max_min}
  \max\min_{\forall x_k } \! \frac{ M^{\mathrm{C}}\! \left\|\! \sum\limits_{k=1}^{K}\! x_k  (\bm{h}^{\mathrm{C}}_{k} )^H\bm{\phi}_{k,1} \!+\! (1\!-\! x_k)  (\bm{h}^{\mathrm{C}}_{k} )^H\bm{\phi}_{k,2}\right\|^2\!\!P^{\mathrm{T}}}{\sigma^2},
\end{equation}
where  $x_k \in \{0,1\}$ denotes the beamforming vector selection for the $k$-th RIS row.

The two objectives above ensure that the RIS introduces significant fluctuations while maintaining high communication performance. However, the communication objective defined in equation~\eqref{eq:max_min} is highly complex, which poses challenges for subsequent optimization and beamforming design. To address this, we construct a surrogate objective that retains the core design intent but is more tractable for optimization.
Specifically, the original objective in equation~\eqref{eq:max_min} involves maximizing the squared magnitude of a sum of complex-valued signals. Intuitively, this is achieved when the individual complex components are phase-aligned and have large amplitudes. Based on this insight, we define $\varphi$ as the phase of the aggregated signal, i.e., $\varphi = \angle(\sum_{k=1}^{K}  (\bm{h}^{\mathrm{C}}_{k} )^H\bm{\phi}_{k} )$. As shown in Fig.~\ref{fig:opt}(b), we then align each beamformed signal to this phase and seek to maximize the following minimum real component across the two beamforming vectors for each RIS row:
\begin{equation}
    \max~\min_{i\in\{1,2\}}  \mathcal{R}\{ (\bm{h}^{\mathrm{C}}_{k} )^H\bm{\phi}_{k,i} e^{-j\varphi }   \}.
\end{equation}

Building on the above analysis, we formulate a beamforming optimization problem by jointly considering communication performance and privacy preservation, as follows:
\begin{subequations}\label{pb:beamforming}
\begin{eqnarray}
    & \max\limits_{ \bm{\phi}_{k,i}, \varphi} &  \sum_{k=1} ^K \left( \omega_1 \left\| (\bm{h}^\mathrm{S}_k)^H \bm{\phi}_{k,1} \right\|^2 + \omega_1\left\| (\bm{h}^\mathrm{S}_k)^H \bm{\phi}_{k,2} \right\|^2 \right. \nonumber\\ 
    & & ~~~~~~~~- \omega_2 \left\| (\bm{h}^\mathrm{S}_k)^H \bm{\phi}_{k,1} + (\bm{h}^\mathrm{S}_k)^H \bm{\phi}_{k,2} \right\|^2  \nonumber\\ 
    & & ~~~~~~~~ +\omega_3\min_{i\in\{1,2\}} \left.  \mathcal{R}\{ (\bm{h}^{\mathrm{C}}_{k} )^H\bm{\phi}_{k,i} e^{-j\varphi } \}   \right), ~~~~~~~~\\
    & \text{s.t.} & | \bm{\phi}_{k,i} [n]| = 1,~\forall k,i,n, \label{con:1}
\end{eqnarray}
\end{subequations}
where $\omega_1$, $\omega_2$, and $\omega_3$ are weighting factors that balance privacy preservation and communication performance. Constraint~\eqref{con:1} enforces the unit-modulus condition on each element of the RIS beamforming vectors.
\revhyh{Specifically, $\omega_1$ controls the sensing-direction gain, $\omega_2$ controls the discrepancy between the two sensing-direction responses, and $\omega_3$ controls the communication-direction stability. A larger $\omega_3$ prioritizes communication performance but may reduce the achievable sensing gain and perturbation strength, while a smaller $\omega_3$ has the opposite effect. In practice, we first tune $\omega_2$ to ensure sufficient perturbation, and then choose $\omega_3$ according to the required communication-direction gain and stability.}

\subsection{Beamforming Design}
To solve problem~\eqref{pb:beamforming}, we adopt a BCD method, which partitions the variables into multiple blocks and updates them iteratively in a cyclic manner. The update procedure consists of the following steps: 1) with all other variables fixed, each element of the beamforming vector $\bm{\phi}_{k,1}$ is updated sequentially; 2) Similarly, each element of $\bm{\phi}_{k,2}$ is updated while keeping the remaining variables fixed; 3) given the beamforming vectors $\bm{\phi}_{k,1}$ and $\bm{\phi}_{k,2}$, the communication-phase variable $\varphi$ is updated accordingly.

In Step 1, we sequentially optimize each element $\bm{\phi}_{k,1}[n]$ for all $k$ and $n$. Among the four components in the original objective function, three terms are dependent on $\bm{\phi}_{k,1}[n]$: the first, third, and fourth. To formulate the subproblem for $\bm{\phi}_{k,1}[n]$, we analyze each of these terms individually as follows:

\noindent $\bullet$ The first term can be rewritten as: $\omega_1 \left( |\bm{h}^\mathrm{S}_k[n]|^2 + 2\mathcal{R}\{ \beta_{k,n,1}^H (\bm{h}^\mathrm{S}_k[n])^H \bm{\phi}_{k,1}[n] + |\beta_{k,n,1}|^2 \} \right)  $, where $\beta_{k,n,1} = \sum_{n'\neq n} (\bm{h}^\mathrm{S}_k[n'])^H \bm{\phi}_{k,1}[n']$;

\noindent $\bullet$ The second term can be rewritten as: $-\omega_2 \left( |\bm{h}^\mathrm{S}_k[n]|^2 + 2\mathcal{R}\{ \beta_{k,n,2}^H (\bm{h}^\mathrm{S}_k[n])^H \bm{\phi}_{k,1}[n] + |\beta_{k,n,2}|^2 \} \right)  $, where  $\beta_{k,n,2} = \sum_{n'\neq n} (\bm{h}^\mathrm{S}_k[n'])^H \bm{\phi}_{k,1}[n'] +(\bm{h}^\mathrm{S}_k)^H \bm{\phi}_{k,2} $; 

\noindent $\bullet$ The third term can be rewritten as: $\omega_3 \min\{\mathcal{R}\{ \eta_{k,n,3}\bm{\phi}_{k,1}[n]\}  + \beta_{k,n,3},\beta_{k,n,4} \}$, where $\eta_{k,n,3} = e^{-j\varphi }(\bm{h}^\mathrm{C}_k[n])^H$, $\beta_{k,n,3} =\mathcal{R}\{\sum_{n'\neq n} (\bm{h}^\mathrm{C}_k[n'])^H \bm{\phi}_{k,1}[n'] e^{-j\varphi } \}$, and $\beta_{k,n,4} = \mathcal{R}\{ (\bm{h}^\mathrm{C}_k)^H \bm{\phi}_{k,2} e^{-j\varphi }\} $.
\revhyh{Detailed derivations can be found in the supplemental material.}

Based on the above reformulation, and after omitting constant terms that are independent of $\bm{\phi}_{k,1}[n]$, the subproblem for optimizing $\bm{\phi}_{k,1}[n]$ can be expressed as:
\begin{subequations}\label{spb:phi}
\begin{eqnarray}
   &\max\limits_{\bm{\phi}_{k,1}[n]} & \omega_3 \min\{\mathcal{R}\{ \eta_{k,n,3}\bm{\phi}_{k,1}[n]\}  + \beta_{k,n,3},\beta_{k,n,4} \}   \nonumber \\
   & &+  \mathcal{R} \{ \eta_{k,n,1} \bm{\phi}_{k,1}[n] \}, \\
   & \text{s.t.} & | \bm{\phi}_{k,1} [n]| = 1, 
\end{eqnarray}
\end{subequations}
where $ \eta_{k,n,1}= 2(\omega_1 \beta_{k,n,1}^H (\bm{h}^\mathrm{S}_k[n])^H  - \omega_2  \beta_{k,n,2}^H (\bm{h}^\mathrm{S}_k[n])^H )$. Then, the optimal solution is given in the following theorem.
\begin{thm} \label{thm:phi}
The optimal solution to problem~\eqref{spb:phi} falls into one of two cases:

\noindent $\bullet$  If there are two distinct phase angles $\bm{\phi}_{k,1}[n]$ (denoted by $\bm{\phi}_{k,1}^{(1)}[n]$ and $\bm{\phi}_{k,1}^{(2)}[n]$) satisfying $\mathcal{R}\{\eta_{k,n,3}\bm{\phi}_{k,1}[n]\}  + \beta_{k,n,3}=\beta_{k,n,4}$, the optimal solution must be selected from the following four candidates: $\bm{\phi}_{k,1}^{(1)}[n]$, $\bm{\phi}_{k,1}^{(2)}[n]$,  $e^{-j\angle (\eta_{k,n,1}+\omega_3\eta_{k,n,3}) }$, and $e^{-j\angle (\eta_{k,n,1}) }$. The final solution is chosen by evaluating the objective function at these candidates and selecting the one with the maximum value.

\noindent $\bullet$  If such $\bm{\phi}_{k,1}^{(1)}[n]$ and $\bm{\phi}_{k,1}^{(2)}[n]$ do not exist, it implies that one term in the $\min$ always dominates the other. Specifically, if $\mathcal{R}\{\eta_{k,n,3}\bm{\phi}_{k,1}[n]\}  + \beta_{k,n,3}\le \beta_{k,n,4}$ holds for all $\bm{\phi}_{k,1}[n]$, then the optimal solution is given by $e^{-j\angle (\eta_{k,n,1}+\omega_3\eta_{k,n,3}) }$; otherwise, the optimal solution is $e^{-j\angle (\eta_{k,n,1}) }$.

\begin{IEEEproof}
    Please refer to the supplemental material.
\end{IEEEproof}
\end{thm}

In Step 2, we sequentially optimize $\bm{\phi}_{k,2}[n]$ for all $k$ and $n$. Since this step closely resembles Step 1, we omit the detailed derivations for brevity. In Step 3, we optimize the phase variable $\varphi$. The corresponding optimization problem is formulated as:
\begin{equation}
    \max\limits_{  \varphi} ~ \sum\nolimits_{k=1} ^K  \min_{i\in\{1,2\}}  \mathcal{R}\{ (\bm{h}^{\mathrm{C}}_{k} )^H\bm{\phi}_{k,i} e^{-j\varphi } \}.
\end{equation}
\revhyh{The main challenge here arises from the non-smoothness of the objective function due to the $\min$ operator, and optimizing this objective requires determining, for each $k$, which of the two terms is smaller. To this end, we identify the switching points by equating the two terms. Specifically, for each $k$, the corresponding switching points are given by
\begin{equation}
    \varphi_k^\mathrm{sw} = -\angle((\bm{h}^{\mathrm{C}}_{k} )^H\bm{\phi}_{k,1} -(\bm{h}^{\mathrm{C}}_{k} )^H\bm{\phi}_{k,2} ) + \frac{\pi}{2}+i\pi,~i\in \mathbb{Z},
\end{equation}
at which $ \mathcal{R}\{ (\bm{h}^{\mathrm{C}}_{k} )^H\bm{\phi}_{k,1} e^{-j\varphi } \} = \mathcal{R}\{ (\bm{h}^{\mathrm{C}}_{k} )^H\bm{\phi}_{k,2} e^{-j\varphi } \}$.}
Restricting $\varphi_k^\mathrm{sw}$ to the interval $[0, 2\pi]$ yields a sorted set of at most $2K$ distinct breakpoints. These divide the domain into subintervals where, for each $k$, the index $i_k \in \{1,2\}$ minimizing the inner expression remains fixed. 
\revhyh{Therefore, the $\min$ operation can be removed, and the objective becomes smooth. The corresponding subproblem can be rewritten as
\begin{equation} \label{spb:varphi}
    \max~\mathcal{R}\{ \sum\nolimits_{k=1}^K(\bm{h}^{\mathrm{C}}_{k} )^H\bm{\phi}_{k,i_k}  e^{-j\varphi }\}.
\end{equation}}
For each subinterval, we compute the optimal $\varphi$ by solving problem~\eqref{spb:varphi}, and also evaluate the objective function at the corresponding interval boundaries. The final solution is obtained by selecting the $\varphi$ that yields the maximum objective value among all candidates.

{\begin{algorithm}[t]
\caption{The overall BCD-based beamforming design algorithm to problem~\eqref{pb:beamforming}.}\label{alg:beamforming}
\DontPrintSemicolon
Define the tolerance of accuracy $\delta$. Initialize the algorithm with a feasible point. Set $l=0$ and the maximum iteration number $L_{\max}$;\;
\Repeat{The decrease of the objective function is less than $\delta$ or the maximum number of iterations is reached, i.e., $l\ge L_{\max}$}{
    Update $\bm{\phi}_{k,1}[n]$, $\forall n,k$ according to Step 1;\;
    Update $\bm{\phi}_{k,2}[n]$, $\forall n,k$ according to Step 2;\;
    Update $\varphi$ according to Step 3;\;
    Update the iteration number: $l\leftarrow l+1$;\;
    }
\end{algorithm}
}

With these three steps, we iteratively solve problem~\eqref{pb:beamforming} with the BCD framework. The overall BCD-based beamforming design algorithm is summarized in Algorithm~\ref{alg:beamforming}, and its computational complexity can be analyzed in the following. 
In Step 1, solving problem~\eqref{spb:phi} for each element $\bm{\phi}_{k,1}[n]$ requires $\mathcal{O}(N)$ operations. Since this update is performed for all $n$ and $k$, the total complexity of Step 1 is $\mathcal{O}(N^2K)$.
Similarly, Step 2 has the same complexity, i.e., $\mathcal{O}(N^2K)$.
In Step 3, solving each instance of problem~\eqref{spb:varphi} involves $\mathcal{O}(K)$ operations. As there are up to $\mathcal{O}(K)$ subintervals to evaluate (due to the at most $2K$ switching points), the total complexity of this step is $\mathcal{O}(K^2)$. In summary, the total computational complexity of Algorithm~\ref{alg:beamforming} is $\mathcal{O}\left( I_{\max}(2N^2K+K^2) \right)$ where $I_{\max}$ denotes the maximum number of BCD iterations. Moreover, regarding convergence, each subproblem in Algorithm~\ref{alg:beamforming} is solved exactly and optimally within the BCD framework. Therefore, according to Proposition~2.7.1 in~\cite{bertsekas1999nonlinear} for the convergence of the BCD framework, Algorithm~\ref{alg:beamforming} is guaranteed to converge to a Karush-Kuhn-Tucker (KKT) point of problem~\eqref{pb:beamforming}, i.e., a stationary point satisfying the KKT conditions.

\subsection{Compatibility with 1-bit RIS}

Given that most practical RIS hardware supports only 1-bit phase resolution, i.e., $\bm{\phi}_{k,i}[n]\in\{-1,1\}$, we extend our proposed beamforming design algorithm to accommodate such constraints in this section.
Under this setting, the optimization problem becomes an instance of integer programming, which is typically NP-hard. To address this challenge, we relax the binary constraint by treating $\bm{\phi}_{k,i}[n]$ as a continuous real-valued variable constrained to  $[-1,1]$. To attract the solution to converge to valid binary values, we introduce a penalty term into the objective function: $\rho((\bm{\phi}_{k,i}[n])^2-1)$, where $\rho$ is a tunable penalty factor that is adaptively adjusted during the iterative optimization process. Notably, when $\bm{\phi}_{k,i}[n]$ is $\pm 1$, the penalty term is zero; otherwise, it becomes negative, thereby lowering the overall objective value and discouraging infeasible solutions. This strategy effectively guides the optimization toward the desired binary outputs.

With this modification, the update rule for $\bm{\phi}_{k,i}[n]$ must be adjusted accordingly. Taking $\bm{\phi}_{k,1}[n]$ as an example, the corresponding subproblem becomes:
\begin{subequations}\label{spb:phi2}
\begin{eqnarray}
   \!\!\!&\max\limits_{{\small \bm{\phi}_{k,1}[n] \in \mathbb{R}}} \!\!\!& (\omega_1\!-\!\omega_2)  |\bm{h}^\mathrm{S}_k[n]|^2 (\bm{\phi}_{k,1}[n])^2 + \rho((\bm{\phi}_{k,1}[n])^2-1)  \nonumber \\
    \!\!\!& &
   +\omega_3 \min\{\mathcal{R}\{ \eta_{k,n,3}\}\bm{\phi}_{k,1}[n] \!+\! \beta_{k,n,3},\beta_{k,n,4} \} \nonumber\\
    \!\!\!& &+ \mathcal{R}\{ \eta_{k,n,1}\} \bm{\phi}_{k,1}[n],  \\
    \!\!\! & \text{s.t.} & -1\le \bm{\phi}_{k,1}[n]\le 1. 
\end{eqnarray}
\end{subequations}
The above subproblem is a quadratic optimization problem, which can be efficiently solved using standard methods. Due to space limitations, we omit the detailed derivation here.
\revhyh{Now, we obtain an extended version of our algorithm, as shown in Algorithm~\ref{alg:beamforming_1bit}. This algorithm follows a double-loop structure: the outer loop updates the penalty factor $\rho$, while the inner loop applies the BCD method to optimize the revised objective function. The inner loop is guaranteed to converge to a stable stationary solution under the BCD updates, and the increasing penalty factor in the outer loop forces the phase constraint to be satisfied; hence, at convergence, the final phase of each reflecting element must take either $\pi$ or $0$.}

{\begin{algorithm}[t]
\caption{\revhyh{The overall penalty-based beamforming design algorithm for 1-bit RIS.}}\label{alg:beamforming_1bit}
\DontPrintSemicolon
Define the tolerance of accuracy $\delta$ and the penalty factor $\rho$. Initialize the algorithm with a feasible point. Set $l=0$ and the maximum iteration number $L_{\max}$;\;
\Repeat{$\sum_{k,i,n}|(\bm{\phi}_{k,i}[n])^2-1|<\delta$}{
\Repeat{convergence}{
    Update $\bm{\phi}_{k,1}[n]$, $\forall n,k$ by solving problem~\eqref{spb:phi2};\;
    Update $\bm{\phi}_{k,2}[n]$, $\forall n,k$, similarly;\;
    Update $\varphi$ according to Step 3;\;
    Update the iteration number: $l\leftarrow l+1$;\;
    }
Update the penalty factor: $\rho\leftarrow 3\rho$;\;
}
\end{algorithm}
}

\section{The Design of \name} \label{sec:design}

\begin{figure}[t]
    \centering
    \includegraphics[width=0.88\linewidth]{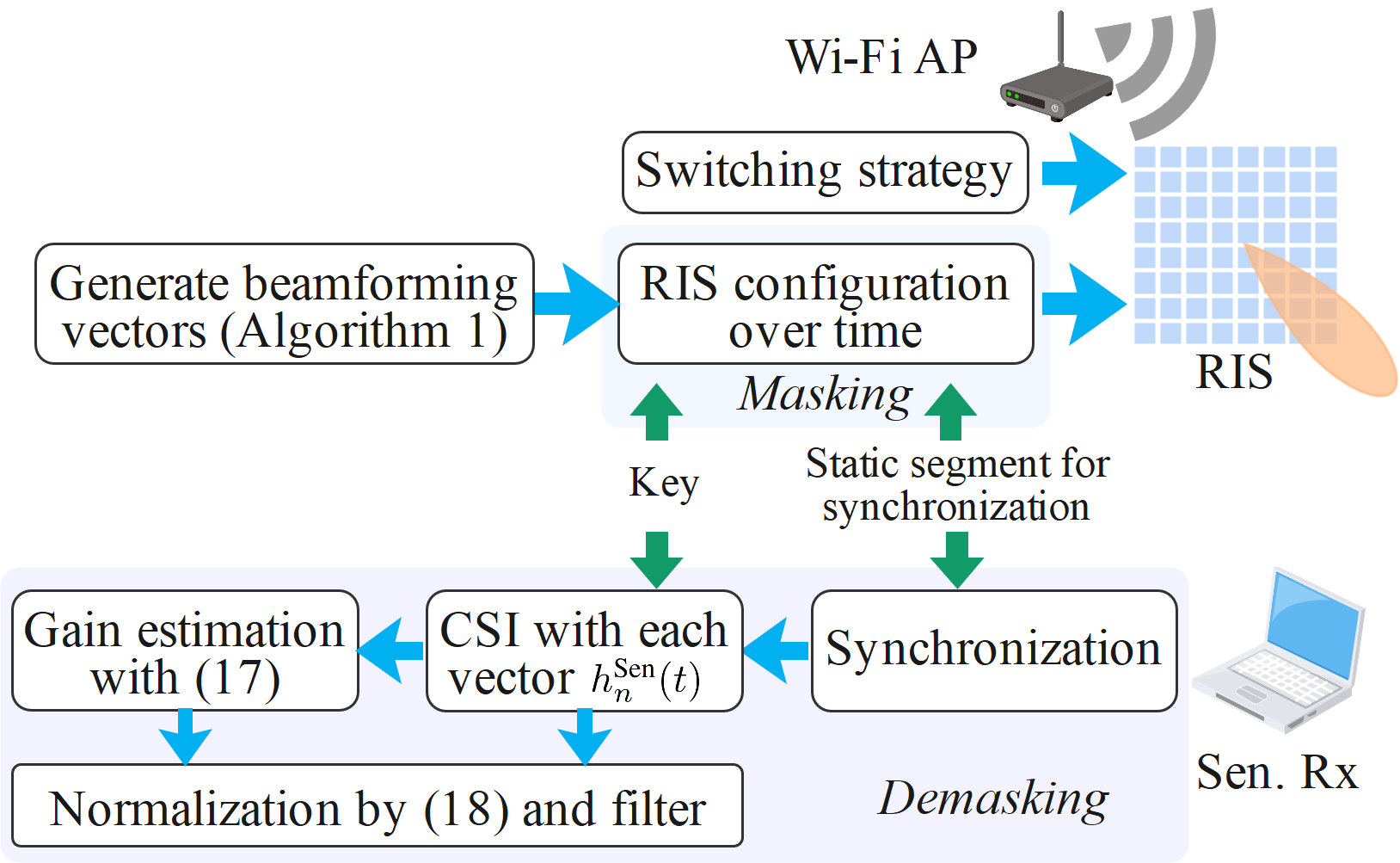}
    \vspace{-1ex}
    \caption{{Overview of \name.}}
    \label{fig:overview}
    \vspace{-1ex}
\end{figure}

After finalizing the beamforming design, this section presents the workflow of \name, as shown in Fig.~\ref{fig:overview}.
The Tx first receives access requests from both the legitimate communication Rx and sensing Rx. Upon receiving the requests, the Tx estimates its channels to both the communication Rx and the sensing Rx, which are used to obtain $\vartheta^{\mathrm{C}}$ and $\vartheta^{\mathrm{S}}$. This can be achieved using existing RIS-based channel estimation and localization algorithms~\cite{zheng2020intelligent,zheng2019intelligent,li2023riscan,fan2025sense}. Then, the Tx executes Algorithm~\ref{alg:beamforming} to determine the RIS beamforming vectors. 
{Simultaneously, a digital key is securely shared between the Tx and the legitimate sensing Rx\footnote{\revhyh{We can reuse existing Wi-Fi authentication and encryption mechanisms to share a fresh digital key between the Tx and the legitimate Rx after association. This key is implemented as a random seed for generating the RIS configuration sequence and can be periodically refreshed.}}. Using this key, the Tx applies the time-domain masking method to generate time-varying RIS configurations and transmits packets under the proposed RIS switching strategy, thereby enabling high-performance sensing and communication with \revhyh{effective privacy protection}. At the sensing Rx, the collected CSI is processed using the time-domain demasking procedure, which leverages the shared key to recover the clean CSI, after which standard sensing algorithms are applied for activity recognition.} The workflow of \name involves two key components: 1) RIS switching strategy in Section~\ref{sec:switch}, which addresses the communication disruption caused by RIS beamforming transitions (as discussed in Section~\ref{sssec:comm}); 2) Time-domain masking and demasking method in Section~\ref{subsec:encryption}, which protects sensitive sensing information from potential eavesdroppers while enabling the legitimate sensing Rx to achieve high-performance sensing.

\subsection{RIS Switching Strategy} \label{sec:switch}

\begin{figure}[t]
    \centering
    \setlength{\abovecaptionskip}{6pt}
    \includegraphics[width=0.9\linewidth]{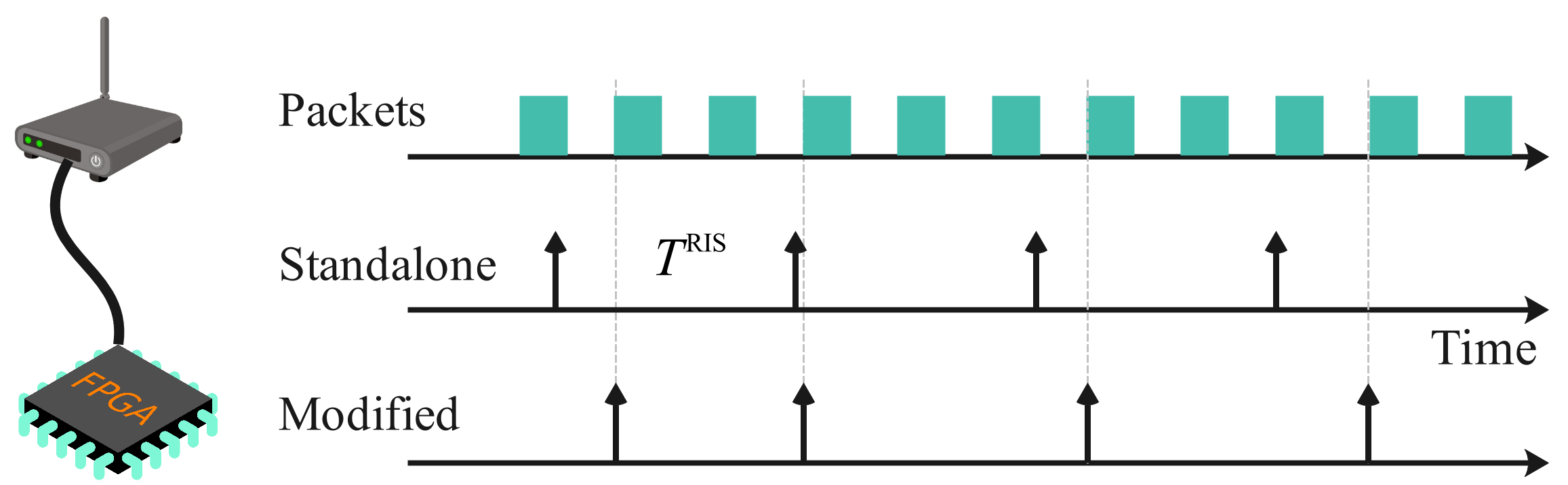}
    \vspace{-0.2em}
    \caption{Timing diagram with proposed strategy.}
    \label{fig:timing_diagram}
    \vspace{-2ex}
\end{figure}

To address the communication disruption caused by RIS phase transitions, we achieve synchronization between the Tx and the RIS control module, e.g., FPGA, via a wired connection. Specifically, let $T^{\mathrm{RIS}}$ denote the configuration switching period for RIS when it works in standalone operation. 
Before transmitting each packet, the Tx sends a trigger signal to the RIS via the wired connection.
Upon receiving this signal, the RIS checks whether it is the first trigger within the current $T^{\mathrm{RIS}}$ period. If so, it updates its beamforming vectors; otherwise, it retains the current configuration. The corresponding timing diagram is shown in Fig.~\ref{fig:timing_diagram}. 
As illustrated, this design ensures that RIS configuration updates do not occur during packet transmission, thereby eliminating the risk of communication disruption. From the Rx's perspective, the RIS still switches approximately once every $T^{\mathrm{RIS}}$, preserving the intended update frequency.

\vspace{-0.4em}
\subsection{Time-Domain Masking and Demasking}
\label{subsec:encryption}
Thus far, we have proposed a BCD-based beamforming design algorithm for the RIS. By randomly selecting a beamforming vector for each row, we construct $N^{\mathrm{R}}$ candidate RIS beamforming configurations. During each interval $T^{\mathrm{RIS}}$, one configuration is randomly activated, thereby introducing temporal fluctuations that obscure sensitive information (such as those illustrated in Fig.~\ref{fig:csi_pre}) and protecting privacy.
The remaining issues lie in enabling the legitimate sensing Rx to accurately extract target-related information. As discussed in Section~\ref{sssec:sen_rx}, the received CSI must be correctly associated with the corresponding RIS configuration and further normalized to eliminate artificial fluctuations. 
To this end, two key questions must be answered:
(1) How can the Rx accurately identify which candidate configuration is activated at each time slot using the shared secret key?
(2) How should the CSI obtained under different RIS configurations be normalized, in order to recover stable and meaningful sensing information?

To address the first question, time synchronization between the RIS and the Rx is essential to ensure that the Rx can correctly map each received CSI sample to its corresponding RIS beamforming vector selection.
\revhyh{To enable synchronization, we embed a predefined RIS configuration within the configuration sequence at a fixed interval, denoted by $T^{\mathrm{sync}}$.} 
Since the Rx cannot directly infer the RIS configuration from raw CSI fluctuations, variations in the configuration over time cannot serve as reliable timing markers. Instead, we adopt a strategy in which the RIS maintains a fixed configuration for a short duration (e.g., across $3T^{\mathrm{RIS}}$). This results in a detectable static segment that the Rx can detect as a synchronization reference.

\begin{figure}[t]
    \centering
    \includegraphics[width=0.85\linewidth]{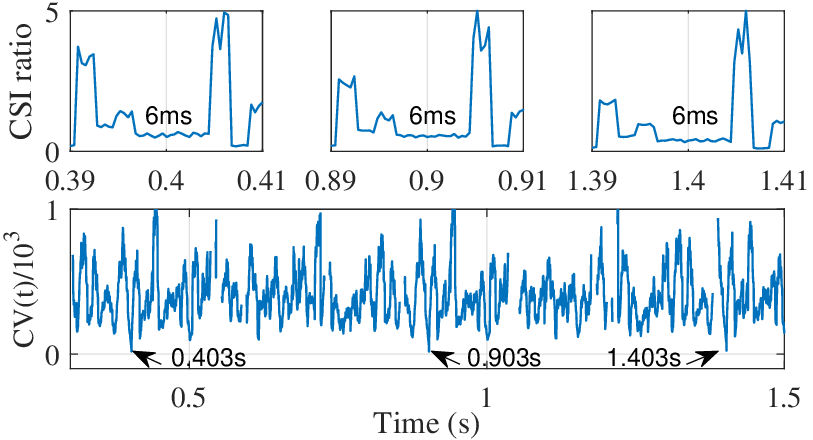}
    \vspace{-2ex}
    \caption{Time synchronization between the Rx and the RIS with $T^{\mathrm{RIS}}$ being 2~\!ms and $T^{\mathrm{sync}}$ being 0.5~\!s.} 
    \label{fig:time_sync}
    \vspace{-1.8em}
\end{figure}

At the Rx side, once CSI is collected, it is compared with CSI samples from the previous $3T^{\mathrm{RIS}}$ window to detect the static segments. To mitigate the influence of time-varying phase distortions and Rx-side interference, we adopt a CSI ratio~\cite{zeng2019farsense} by dividing the CSI values between antenna pairs, thus generating time-series signals that can sensitively reflect the CSI variations.
For our adopted three-antenna WiFi NIC, we compute three such CSI ratio sequences: antenna 1 over antenna 2, antenna 2 over antenna 3, and antenna 3 over antenna 1. These sequences are denoted as $h^{\mathrm{S,R}}_m(t)$, where $m$ indexes the antenna-pair ratio streams.
To determine whether the CSI is in a static state, we apply the coefficient of variation (CV), which measures relative signal fluctuation independent of absolute amplitude. For each subcarrier $f$ and antenna ratio stream $m$, it is defined as:
\begin{equation}
  \mathrm{CV}_{f,m}(t) = \frac{\mathrm{SD}_{\tau\in[t-3T^{\mathrm{RIS}},t]}\{h^{\mathrm{S,R}}_m(\tau) \}}{ |\mathrm{Mean}_{\tau\in[t-3T^{\mathrm{RIS}},t]}\{h^{\mathrm{S,R}}_m(\tau)\}|},
\end{equation}
where $\mathrm{SD}\{\cdot\}$ and $\mathrm{Mean}\{\cdot\}$ represent the standard deviation and mean over time, respectively.
To suppress noise and enhance detection reliability, we aggregate the CV values across all subcarriers and antenna ratio streams:
\begin{equation}
  \overline{\mathrm{CV}}(t) = \sum_f \sum_m \mathrm{CV}_{f,m}(t).
\end{equation}
Since the synchronization selection appears only once in each $T^{\mathrm{sync}}$ interval and lasts for only a few milliseconds, there is guaranteed to be a single, distinct synchronization point within any randomly selected $T^{\mathrm{sync}}$ interval, corresponding to the minimum value of $\overline{\mathrm{CV}}(t)$. 
To further improve accuracy, the Rx can apply linear least-squares estimation over multiple synchronization points.
Fig.~\ref{fig:time_sync} illustrates this process, where the interval between synchronization codes is set to 0.5 seconds. As shown, each 0.5-second window contains a unique global minimum in the aggregated CV curve, corresponding precisely to the end of the synchronization segment.
By identifying these minima, the RIS and Rx can establish accurate time alignment, ensuring a correct mapping between CSI samples and the RIS configurations using the shared key.

After addressing the synchronization issue, the received CSI can be accurately mapped to the $N^{\mathrm{R}}$ RIS configurations using the shared key, denoted as $h_{n}^{\text{Sen}}(t),n=1,\cdots,N^{\mathrm{R}}$. As indicated in sensing channel model~\eqref{eq:channel_sen}, the sensing gain (i.e., transmit power toward the sensing direction) differs across RIS configurations, and removing artificial perturbations essentially requires eliminating this gain. However, this gain cannot be directly obtained at the Rx side. Therefore, we need to estimate the gain for each configuration. To achieve this, we first eliminate the impact of static paths unrelated to the target. Specifically, we calculate the temporal mean of $h_{n}^{\text{Sen}}(t)$ and subtract it from the raw sequence, i.e.,
\begin{equation}
\bar{h}_{n}^{\text{Sen}}(t) = h_{n}^{\text{Sen}}(t) - \underset{\textrm{over }t}{\textrm{Mean}}\{h_{n}^{\text{Sen}}(t)\},~\forall n.
\end{equation}
Since obtaining the absolute gains of all configurations is challenging, we instead focus on their relative gains. In this process, one configuration (e.g., the first one) is selected as the reference, and all other configurations are then normalized relative to this reference. 
{To estimate the gain, we leverage the fact that the CSI remains nearly constant within each channel coherence interval.}
We traverse the CSI sequence to identify adjacent packets whose inter-packet intervals are below a predefined threshold and that belong to different configurations. By dividing the CSI values of these adjacent packets, we obtain an estimate of their relative gain, denoted as $w_{n_1,n_2}$ for the $n_1$-th and $n_2$-th configurations. Multiple such estimates are collected, and their average value $\bar{w}_{n_1,n_2}$ is taken to mitigate noise.  
This process yields a relative gain matrix $\mathbf{W} = [\bar{w}_{n_1,n_2}] \in \mathbb{R}^{N^{R}\times N^{R}}$ for different RIS configurations, and its diagonal elements are all ones.\footnote{{We do not require all RIS configurations to appear within a single coherence interval. For each relative gain estimate, it is sufficient that a coherence-time segment contains two RIS configurations, and as the RIS configuration sequence varies over time, we naturally accumulate a sufficiently rich set of such pairs to construct the complete matrix $\mathbf{W}$.} The successful acquisition of $\mathbf{W}$ is attributed to our deliberate restriction to a small configuration set.}
Denoting the gain of the $n$-th configuration relative to the first as $g_n$, we can formulate the following problem to estimate $g_n$:
\begin{subequations}
\begin{eqnarray}
 &\min\limits_{\{g_n\}} & \sum_{n_1=1}^{N^{R}} \sum_{n_2=n_1+1}^{N^{R}} \big| g_{n_1} - \bar{w}_{n_1,n_2} g_{n_2} \big|^2,\\
 & \text{s.t.} & g_1 = 1.
\end{eqnarray}
\end{subequations}
The goal is to find a set of gains $\{g_n\}$ that best fit the relative gain matrix $\mathbf{W}$, with the reference configuration fixed to $g_1 = 1$. This problem is quadratic and can be efficiently solved via convex optimization. Once the relative gains $\{g_n\}$ are obtained, the CSI sequences are normalized accordingly:
\begin{equation}
    \hat{h}_{n}^{\text{Sen}}(t) = \bar{h}_{n}^{\text{Sen}}(t) /g_n, ~\forall n.
\end{equation}
$\hat{h}_{n}^{\text{Sen}}(t)$ from all RIS configurations is further combined into one CSI sequence in chronological order.
Finally, we apply a low-pass filter to the demasked CSI to further suppress noise. The filtered signals, as illustrated in Fig.~\ref{fig:time_fre}, exhibit clear temporal patterns that encode meaningful information, demonstrating the effectiveness of the proposed method. The processed CSI across antennas and subcarriers is then fed into sensing algorithms for downstream sensing tasks.

\begin{figure}[t]
    \centering
    \subfloat[With \name]{
        \centering
        \includegraphics[width=0.45\linewidth]{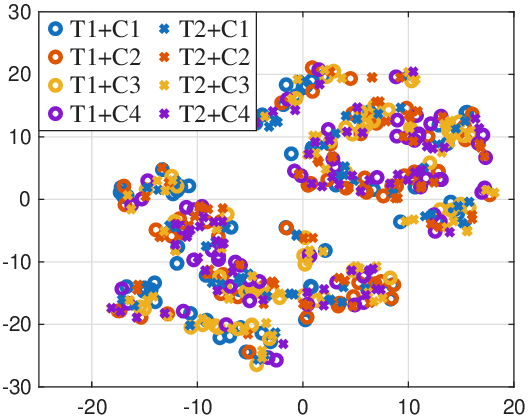}
        \label{fig:tsne}
    }
    \subfloat[Without RIS]{
        \centering
        \includegraphics[width=0.45\linewidth]{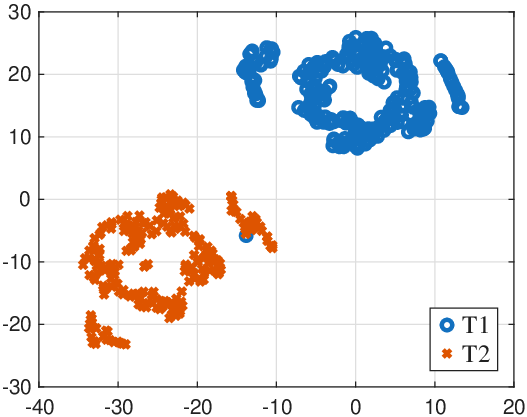}
        \label{fig:tsne_bl}
    }
    \vspace{-1ex}
    \caption{{CSI distribution with t-SNE. Here, ``T1+C1'' means the CSI of gesture 1 under RIS configuration 1.}}\label{fig:tsne_R1}
   \vspace{-3ex}
\end{figure}

\subsection{Security Analysis}

We consider two major threats: (i) an attacker attempting to infer the RIS configuration from CSI and replicate the demasking method, and (ii) an attacker positioning itself arbitrarily and applying passive beamforming to suppress RIS-induced perturbations.

For the first threat, although the RIS configuration set is small, distinguishing RIS configurations from CSI is infeasible in dynamic sensing scenarios. The CSI variations introduced by the RIS are entangled with those caused by human motion, causing samples from different RIS configurations to collapse into overlapping regions. \revhyh{As illustrated by the t-SNE visualization in Fig.~\ref{fig:tsne}, CSI samples under different RIS configurations become completely intermixed, preventing an attacker from identifying state-specific clusters or forming the valid CSI pairs needed for relative-gain estimation. In addition, compared with Fig.~\ref{fig:tsne_bl}, where CSI samples from different gestures form clear and separable clusters without RIS, applying \name causes these clusters to collapse into overlapping regions. This demonstrates that the RIS-induced perturbations effectively mask gesture-dependent CSI signatures, preventing the attacker from distinguishing RIS configurations and extracting private sensing information.} \revhyh{Moreover, the synchronization segments do not provide a useful leakage channel either, as they last only a few milliseconds and use varying RIS states across different synchronization intervals.}

For the second threat, an attacker might attempt to avoid the RIS-influenced region or apply passive beamforming to suppress the signals from the RIS. However, \revhyh{this strategy has limited effectiveness}.
To extract any sensing information, the attacker must rely on the signals reflected from the target. Since \name injects perturbations precisely toward the target direction, these perturbations are inevitably embedded in the target-reflected components that the attacker observes. Any attempt to spatially filter out the perturbations would simultaneously suppress the target reflections, thereby removing the very information the attacker aims to obtain. Consequently, neither positional choices nor passive beamforming allows the attacker to recover the target's private information. Our experiment results in Section~\ref{sec:result} will further confirm this.

\section{Prototype and Experiment Setup} \label{sec:experiment}


\subsection{Implementation}

\begin{figure}[t]
    \centering
    \setlength{\abovecaptionskip}{6pt}
    \includegraphics[width=0.98\linewidth]{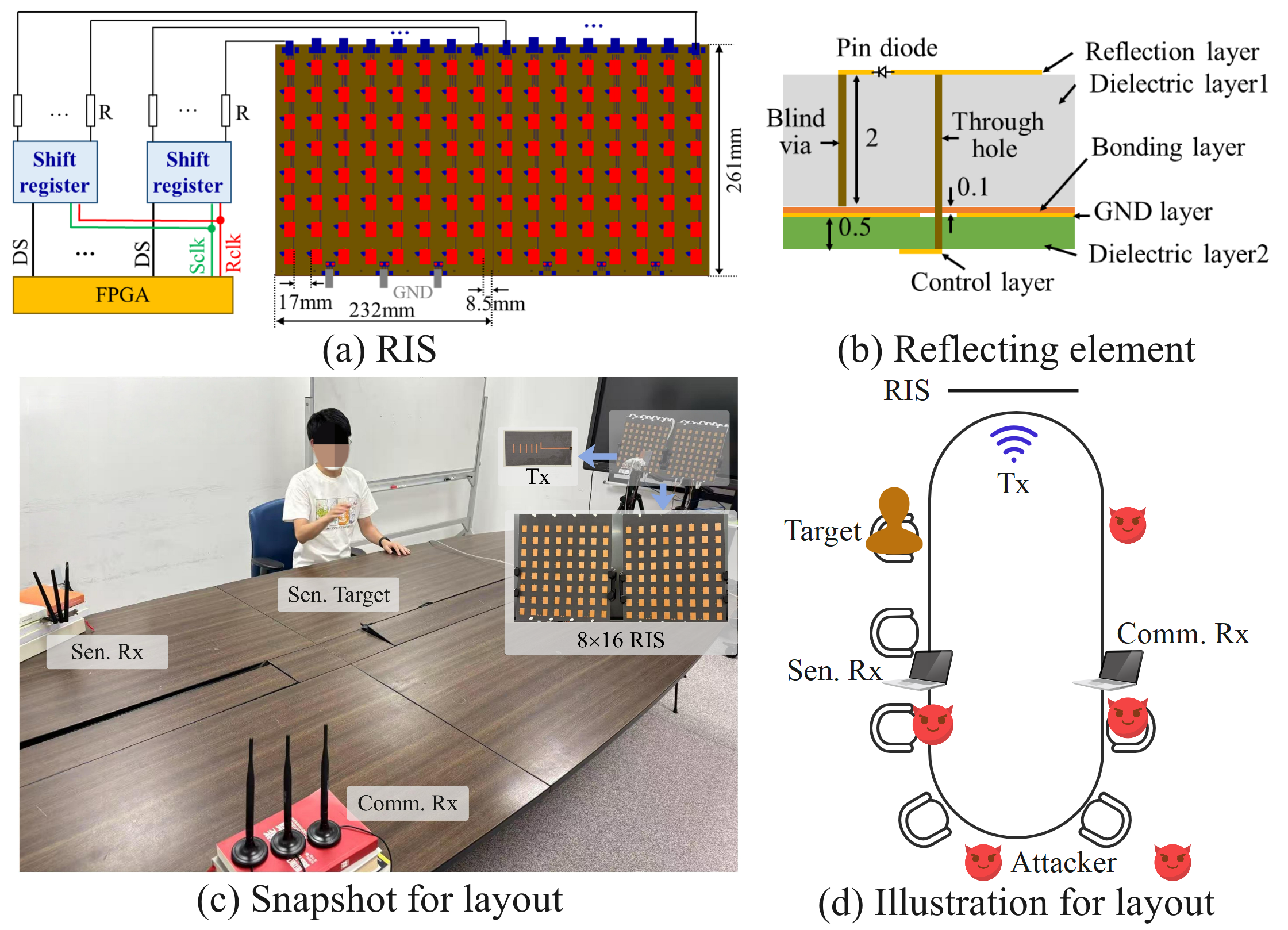}
    \vspace{-2ex}
    \caption{\revhyh{Prototype and experiment setup.}}
    \label{fig:setup}
    \vspace{-1.3em}
\end{figure}

\textbf{RIS Prototype}.
Following~\cite{yang20161}, we develop an RIS prototype consisting of an 8$\times$8 array of elements, forming a planar metasurface with 64 reconfigurable units in total, as shown in Fig.~\ref{fig:setup}. Each element supports 1-bit phase modulation via a surface-mounted MADP-000907-14020x PIN diode, which enables binary phase switching through bias voltage control (0~\!V or 1.35~\!V). The structure of each element adopts a typical design comprising stacked metallic and dielectric layers. By toggling the bias voltage, the reflection phase can be switched between 0 and $\pi$, allowing discrete control over the reflected wavefront at the target frequency of 5.22~\!GHz.
To manage the 64 elements efficiently, the RIS is controlled by an FPGA module. Due to the limited number of general-purpose I/O ports on commercial FPGAs, we integrate serial-in, parallel-out shift registers (e.g., SN74HC595) into the control circuitry. These registers convert the FPGA's 1-bit serial data stream into 8-bit parallel control signals, enabling the sequential loading of configuration bits and simultaneous phase state updates across all elements. \revhyh{Moreover, existing prototype studies~\cite{pei2021ris} indicate that optimized low-power hardware can reduce total RIS power consumption to the sub-watt level.}

\textbf{System Implementation}. 
\name consists of a Tx, a sensing Rx, a communication Rx, and an RIS controlled by an FPGA. The Tx and both Rxs are implemented using mini PCs equipped with Intel 5300 NICs.
To emulate a low-cost IoT device, the transmitter is limited to a single transmit antenna, while both the sensing and communication Rxs are equipped with three antennas each.
The RIS is composed of two 8$\times$8 panels arranged to form a single 8$\times$16 array, as shown in Fig.~\ref{fig:setup}. To ensure that the transmitted signal passes entirely through the RIS, the Tx is equipped with a directional antenna pointed toward the RIS. Meanwhile, Tx is physically connected to the FPGA controller via an RJ45 Ethernet cable.
On the software side, the RIS beamforming vectors obtained using Algorithm~\ref{alg:beamforming} and the time-domain masking method introduced in Section~\ref{subsec:encryption} are implemented on the FPGA using Verilog. The RIS switching strategy detailed in Section~\ref{sec:switch} is jointly implemented in C++ (on the Tx) and Verilog (on the FPGA).
The sensing Rx collects CSI using the PicoScenes~\cite{jiang2021eliminating}. The time-domain demasking method is implemented in MATLAB, while the subsequent sensing algorithms are developed in Python, and model training is conducted on a workstation equipped with an NVIDIA RTX A5000 GPU.

\begin{table}[t]
\centering
\caption{\revhyh{Key evaluation parameters.}}
\label{tab:key_parameters}
\vspace{-1ex}
\small
\begin{tabular}{>{\centering\arraybackslash}m{3cm} >{\raggedright\arraybackslash}m{3.8cm}}
\toprule
\textbf{Parameter} & \textbf{Setting} \\
\midrule
Carrier frequency & 5.22 GHz with bandwidth being 20 MHz \\
Antenna setting & 1 Tx antenna; 3 Rx antennas\\
1-bit RIS size & $8\times16$\\
RIS switching & $T^{\mathrm{RIS}}=2$ ms \\
Synchronization & $T^{\mathrm{sync}}=0.5$ s \\
Packet rate & $\sim$ 500 Hz \\
Sen./Comm. direction & $50^\circ$ / $-20^\circ$ \\
Dataset & 6 volunteers $\times$ 9 gestures $\times$ 50 repetitions\\
\bottomrule
\end{tabular}
\vspace{-4ex}
\end{table}

\subsection{Experiment Setup}

\begin{figure*}[t]
    \centering
    \setlength{\abovecaptionskip}{6pt}
    \subfloat[Convergence]{
        \centering
        \includegraphics[width=0.14\linewidth]{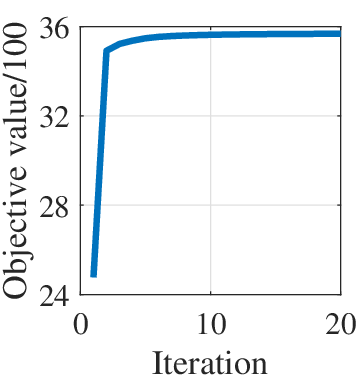}
        \label{fig:conver}
    }
    \subfloat[Proposal (ideal RIS)]{
        \centering
        \includegraphics[width=0.2\linewidth]{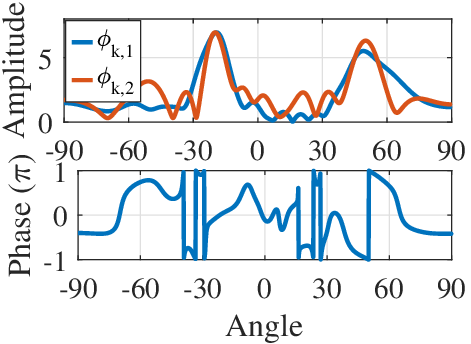}
        \label{fig:beam}
    }
    \subfloat[Proposal (1-bit RIS)]{
        \centering
        \includegraphics[width=0.2\linewidth]{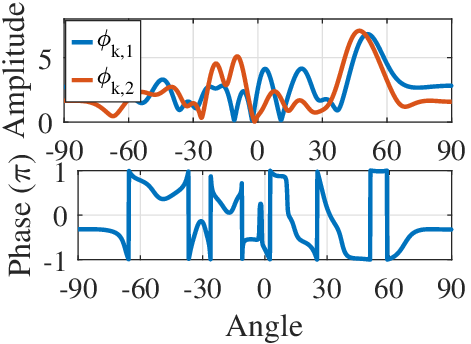}
        \label{fig:beam2}
    }
    \subfloat[Privacy-only]{
        \centering
        \includegraphics[width=0.2\linewidth]{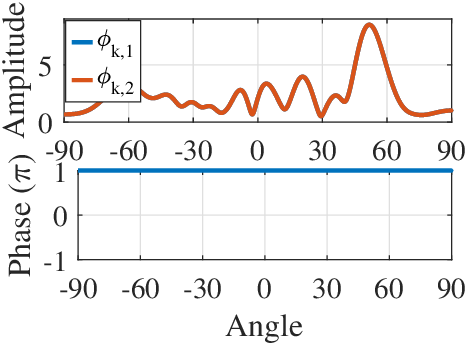}
        \label{fig:beam_sen}
    }
    \subfloat[Communication-only]{
        \centering
        \includegraphics[width=0.2\linewidth]{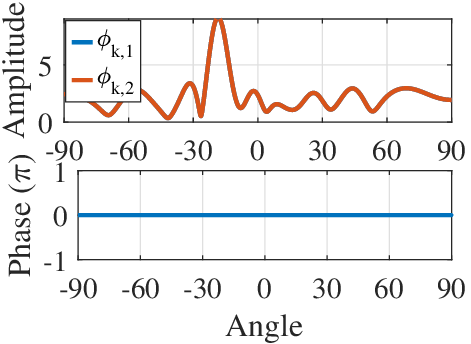}
        \label{fig:beam_com}
    }
    \vspace{-1ex}
    \caption{\revhyh{(a) Convergence behavior of Algorithm~\ref{alg:beamforming} and (b)-(e) beam pattern generated by the first row of the RIS, with the bottom describing the phase difference between two vectors.}}
    \vspace{-1.3em}
\end{figure*}


We begin with a micro-benchmark study to evaluate the beamforming design algorithm proposed in Section~\ref{sec:beamforming}. 
\revhyh{Following existing works~\cite{hu2020reconfigurable}, we obtain the Tx--RIS and RIS--Rx channels using a geometry-assisted free-space propagation model. Specifically, the Tx--RIS channel $\bm{h}^{\mathrm{T}}_k$ is estimated from the distance between the Tx antenna and each RIS element, while the RIS--Rx channel is constructed from the measured Rx--RIS angle under a far-field approximation.}
Then, it is used to generate RIS configurations using the proposed algorithms.
After validating the effectiveness of the beamforming algorithm, we use the obtained beamforming vectors for overall performance evaluation. The experiments for overall performance are conducted in a typical meeting room environment, as illustrated in Fig.~\ref{fig:setup}. All devices operate on the 5.22~\!GHz band with a bandwidth of 20~\!MHz. 
The Tx continuously transmits data packets with a frequency around 500~\!Hz and the period for RIS configuration switching is 2~\!ms.
\revhyh{To fully exploit the RIS beamforming capability, we deploy the RIS close to the Tx, allowing it to capture stronger incident signals.}
The sensing target is located at a direction of 50$^{\circ}$ and the communication Rx is located at a direction of -20$^{\circ}$, both measured relative to the center normal of the RIS. 
\revhyh{By default, the attacker is placed at a distant location to simulate an eavesdropping scenario. To evaluate system robustness, we further test four additional attacker locations, including positions near the legitimate sensing Rx and the legitimate communication Rx.}
We recruit six volunteers (four males and two females) to participate in the experiments. Each participant performs nine distinct gestures: push-pull (PP), slide (SL), up-down (UD), clap (CL), wave (WA), draw circle (DC), draw square (DS), draw zigzag (DZ), and an idle state (IS), where no gesture is performed. Under the default configuration, each participant repeats each gesture 50 times. The resulting dataset is split into training and testing sets with a 7:3 ratio. For other configurations (e.g., varying the RIS size), each gesture is repeated 30 times, with the resulting data used exclusively for testing.
We adopt the gesture classification model in SignFi~\cite{ma2018signfi}.
To evaluate sensing performance for both the legitimate Rx and the attacker, we use classification accuracy as the metric. For communication performance, we report the successful transmission ratio, defined as one minus the packet loss rate. In addition, we include a baseline scenario in which no RIS is deployed and the Tx uses an omnidirectional antenna, to demonstrate the advantages of our proposed design.
All experiments strictly follow the Institutional Review Board guidelines of our institute. \revhyh{The key parameters are listed in Tab.~\ref{tab:key_parameters}.}

\section{Evaluation Result} \label{sec:result}

In this section, we first present a micro-benchmark study, followed by evaluations of \name's privacy protection, sensing accuracy, and communication performance. We then investigate the impact of various system parameters.

\subsection{Micro-benchmark Study}

\begin{figure*}[t]
    \centering
\subfloat[Privacy protection]{
        \centering
        \includegraphics[width=0.19\linewidth]{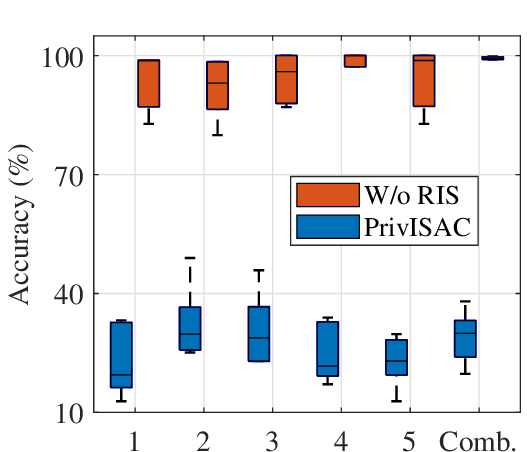}
        \label{fig:acc_attacker1}
    }
\subfloat[Comparison]{
        \centering
        \includegraphics[width=0.15\linewidth]{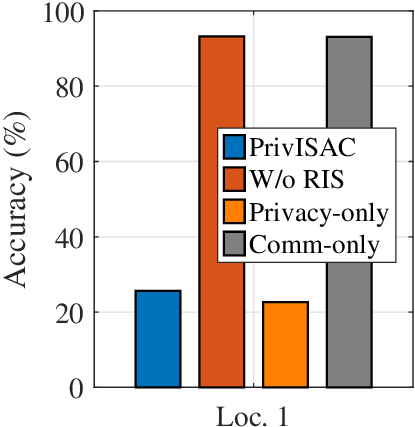}
        \label{fig:acc_attacker}
    }
\subfloat[\name]{
        \centering
        \includegraphics[width=0.19\linewidth]{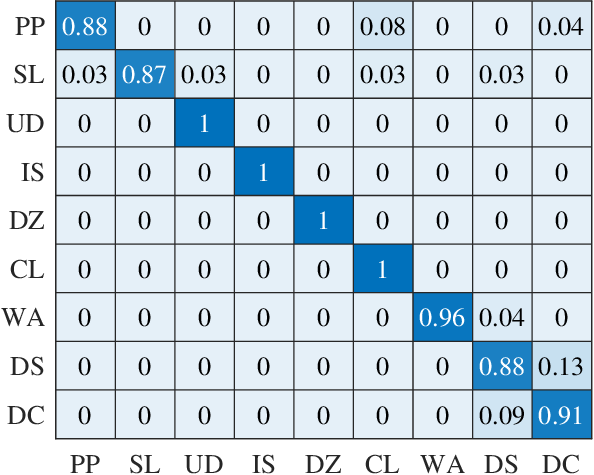}
        \label{fig:rx_ris}
    }
\subfloat[W/o RIS]{
        \centering
        \includegraphics[width=0.19\linewidth]{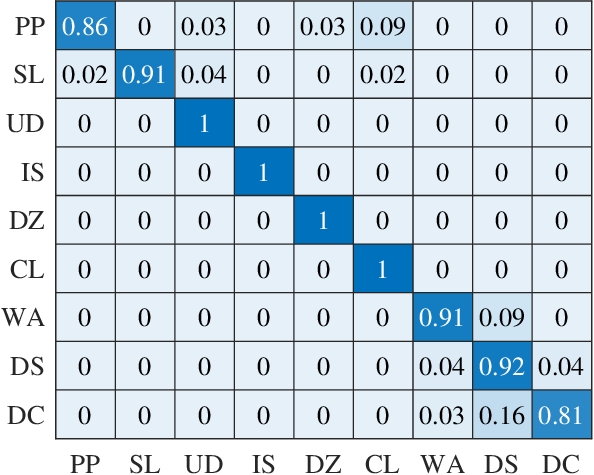}
        \label{fig:rx_bl}
    }
\subfloat[W/o demasking]{
        \centering
        \includegraphics[width=0.19\linewidth]{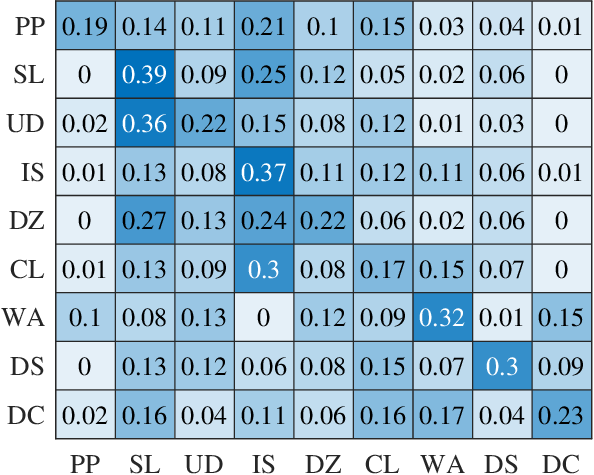}
        \label{fig:rx_ris_wo}
    }
    \vspace{-1ex}
    \caption{\revhyh{(a) Privacy protection at different attacker locations, (b) comparison with baseline schemes at Loc. 1, and (c)--(e) the confusion matrices of the legitimate sensing Rx for (c) \name, (d) without RIS, and (e) \name without demasking (i.e., attacker).}}
    \label{fig:rx_performance}
    \vspace{-3ex}
\end{figure*}

This study aims to validate the effectiveness of the proposed beamforming design.
First, Fig.~\ref{fig:conver} illustrates the convergence behavior of Algorithm~\ref{alg:beamforming}. As shown, the algorithm converges to a stable solution within approximately 10 iterations, demonstrating its fast convergence. Additionally, the objective function exhibits a clear upward trend during the early iterations, highlighting the algorithm's ability to effectively optimize the beamforming vectors.
To further evaluate the design, we visualize the beam pattern generated by the first RIS row in Fig.~\ref{fig:beam}. The top panel shows the signal amplitudes of the two designed beamforming vectors across different angles, while the bottom panel presents their corresponding phase differences. The remaining RIS rows exhibit similar patterns and are thus omitted for brevity.
Notably, both the communication direction ($-20^\circ$) and the sensing direction ($50^\circ$) exhibit strong beamforming gains, confirming that the design supports high-quality communication and sensing. More importantly, in the communication direction, the two beamforming vectors produce nearly identical magnitudes and phases, ensuring that random switching between them does not degrade communication performance.
In contrast, in the sensing direction, the magnitudes remain similar, but the phase difference fluctuates around $\pm\pi$, inducing significant CSI variation. This variation serves to obfuscate the CSI measured at the attacker and enhance privacy protection. 
\revhyh{Furthermore, Fig.~\ref{fig:beam2} shows the beam pattern achieved by the proposed 1-bit RIS design. The 1-bit design still exhibits beam-pattern characteristics similar to the ideal case, verifying the effectiveness and practicality of the proposed 1-bit beamforming design algorithm. Although the beamforming gain is slightly lower than that of the ideal one, it remains sufficient for our system requirements. Moreover, later communication and sensing experiments further confirm that these desired properties remain effective after practical deployment.}

\revhyh{We further compare \name with two single-objective beamforming designs. Fig.~\ref{fig:beam_sen} shows the privacy-only design, where the RIS vectors are optimized only to create sensing-direction perturbations and are randomly switched to mask sensing-related CSI. Although this design produces large phase differences in the sensing direction, its communication-direction response becomes highly unstable across different RIS vectors, explaining its degraded communication performance. Fig.~\ref{fig:beam_com} shows the communication-only design, which uses a fixed RIS configuration to maximize and stabilize the communication-direction gain. However, since it does not introduce time-varying sensing-direction perturbations, it provides limited privacy protection. In comparison, \name explicitly balances these two objectives by maintaining consistent high-gain responses in the communication direction while preserving strong perturbations in the sensing direction.}


\subsection{Overall Performance}

\textbf{Privacy protection}. 
We first evaluate the privacy-preserving capability of \name by measuring the gesture recognition accuracy of an eavesdropping attacker. Fig.~\ref{fig:acc_attacker1} shows the attacker's sensing accuracy at five different locations.
In the baseline scenario without any protection, the attacker achieves an average recognition accuracy of approximately 93~\!\%. After applying \name, the accuracy drops significantly to around 30~\!\%, representing a reduction of over 60~\!\% and highlighting the \revhyh{effective privacy protection} capability of our approach.
\revhyh{The underlying reason is that the RIS is placed between the Tx and the sensing target. Therefore, regardless of the attacker's location, the target-reflected signals received by the attacker are inevitably mixed with RIS-perturbed components. As a result, the observed CSI contains not only target-related information but also artificial variations introduced by the RIS in the sensing direction.}
Crucially, these two components, target-relevant signals and RIS-induced perturbations, are deeply intertwined and indistinguishable from each other in the observed CSI. This makes it fundamentally difficult for an attacker to isolate a meaningful gesture without prior knowledge of the RIS configuration switching pattern.
{To further prove the effectiveness of \name, we aggregate CSI from the five locations and retrain an attack model to fully utilize the multi-view information. One can see that the multi-view attacker achieves nearly 100~\!\% accuracy without \name, but its accuracy drops sharply to 29~\!\% when \name is applied, indicating that even joint multi-point observations cannot recover the target's private sensing information.} 

\revhyh{Moreover, we also compare \name with two representative baselines: communication-only RIS, which uses a fixed configuration to maximize and stabilize the communication-direction gain, and privacy-only RIS, which relies on random switching to introduce sensing-direction perturbations, as shown in Fig.~\ref{fig:acc_attacker}. The results show that both \name and the privacy-only random switching design can effectively reduce the attacker's sensing accuracy, while the communication-only fixed RIS provides limited privacy protection because it does not introduce sufficient time-varying perturbations. The privacy-only design achieves slightly lower attacker accuracy than \name, but it cannot support legitimate sensing and communication, as we discussed in Sec.~\ref{sec:feasibility}.}

\textbf{Sensing performance}. 
Beyond resisting eavesdropping attacks, \name must also ensure high sensing accuracy for the legitimate sensing Rx. To evaluate this, Figs.~\ref{fig:rx_ris} and~\ref{fig:rx_bl} present the confusion matrices of the legitimate Rx under both the proposed \name and a baseline setup without RIS.
In the baseline scenario, the average gesture recognition accuracy reaches 93.3~\!\%. With \name, the accuracy is slightly higher at 94.2~\!\%, suggesting that the masking and demasking method is effective and does not degrade sensing performance. The marginal improvement mainly stems from the additional gain provided by RIS beamforming. 
The RIS beamforming enhances signal power in the sensing direction, improving the effective sensing SNR. It is worth noting that the improvement appears marginal, mainly because the baseline accuracy is already very high.
Moreover, we also observe occasional confusion between the ``drawing a circle'' and ``drawing a square'' gestures. This is likely due to their similar motion trajectories, which make them inherently more difficult to distinguish, even under the baseline.
{Furthermore, we also plot the confusion matrix under \name without proposed demasking method in Fig.~\ref{fig:rx_ris_wo}. As shown, even the legitimate Rx fails to recognize the target's gestures when the demasking method is disabled, which confirms both the necessity and the effectiveness of the proposed demasking method.}



\begin{figure}[t]
    \centering
    \subfloat[Packet duration (ms)]{
        \centering
        \includegraphics[width=0.48\linewidth]{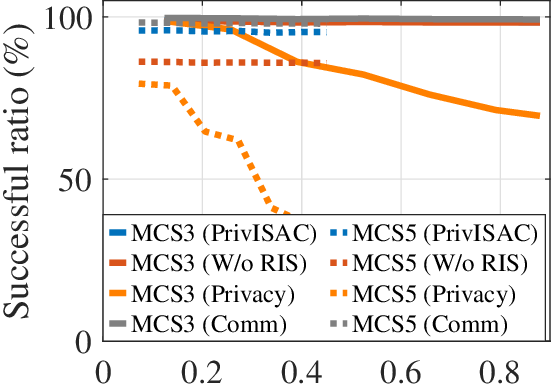}
        \label{fig:fig_per_major}
    }
    \subfloat[RSSI]{
        \centering
        \includegraphics[width=0.48\linewidth]{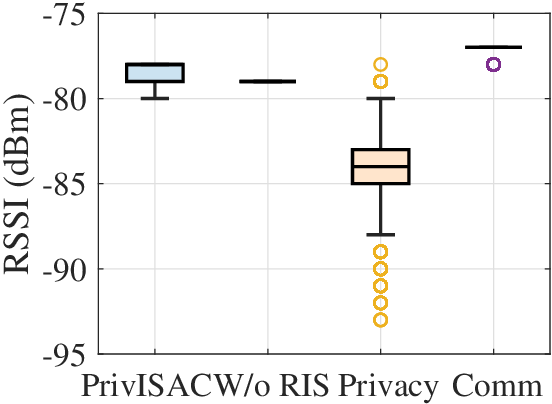}
        \label{fig:fig_rssi_major}
    }
    \vspace{-1ex}
    \caption{\revhyh{(a) Successful ratio under different packet durations and (b) RSSI for four schemes.}}
    \label{fig:fig_per2}
    \vspace{-3ex}
\end{figure}

\textbf{Communication performance}. 
\revhyh{Fig.~\ref{fig:fig_per2} compares the communication performance of \name with the no-RIS baseline, the communication-only RIS scheme, and the privacy-only RIS scheme. As shown in Fig.~\ref{fig:fig_per_major}, \name maintains a high successful transmission ratio under different packet durations and MCS indices, with performance close to the communication-only RIS design. In contrast, the privacy-only random switching design suffers from a clear degradation as packet duration increases, since random RIS transitions cause unstable communication channels. Fig.~\ref{fig:fig_rssi_major} reports the RSSI distribution. Compared with the privacy-only design, \name achieves a much more stable RSSI and remains close to the communication-only design, confirming that the proposed beamforming objective preserves a stable communication-direction response while enabling RIS randomization for privacy protection.}


\begin{figure}[t]
    \centering
    \subfloat[Number of active RIS rows]{
        \centering
        \includegraphics[width=0.48\linewidth]{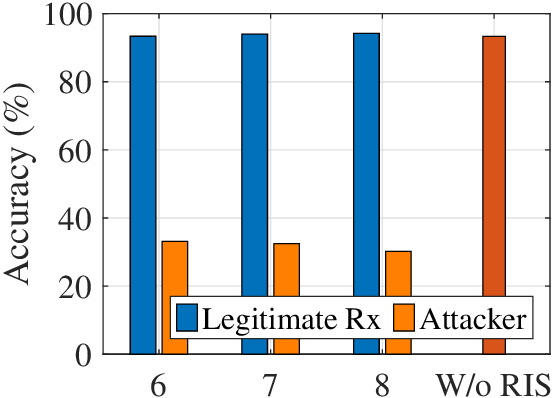}
        \label{fig:acc_ris}
    }
    \subfloat[Number of antennas]{
        \centering
        \includegraphics[width=0.48\linewidth]{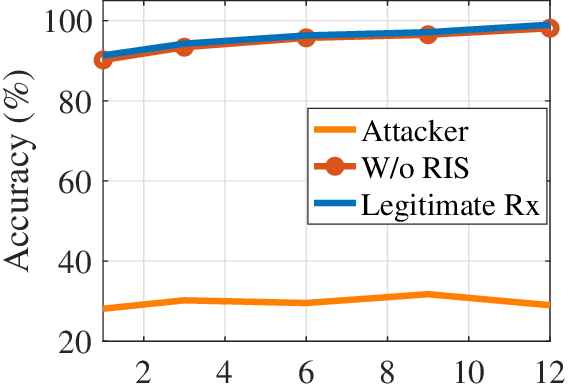}
        \label{fig:acc_anta}
    }
    \vspace{-1ex}
    \caption{{Sensing performance under (a) different numbers of active RIS rows and (b) different numbers of antennas.}}
    \vspace{-3ex}
\end{figure}

\subsection{Impact of Parameters}

\textbf{Impact of RIS size}.
In our proposed system, masking is achieved by randomly selecting one of two beamforming vectors for each row of the RIS. Consequently, the effectiveness of this masking is inherently related to the number of active RIS rows. To evaluate this relationship, we vary the number of activated rows and measure the corresponding sensing performance of both the legitimate Rx and the attacker, as shown in Fig.~\ref{fig:acc_ris}.
As illustrated, the sensing accuracy of the legitimate Rx slightly increases as more RIS rows are activated. This improvement is attributed to enhanced beamforming capability since more active rows allow greater power concentration toward the sensing direction and thus improve the sensing SNR and overall recognition accuracy.
In contrast, the attacker's performance consistently declines with increasing RIS rows. A larger number of rows introduces greater randomness into the measured CSI, thereby enhancing the obfuscation effect and making it more difficult for the attacker to extract meaningful information from the CSI.
Given the relatively low cost and scalability of RIS hardware, configurations such as 8$\times$16 or larger are readily achievable in practice. These results indicate that commodity RIS deployments could offer sufficient capacity to support reliable privacy protection in real-world scenarios.
Additionally, Tab.~\ref{tab:com_rows} demonstrates that the transmission success ratio increases with the number of active RIS rows, owing to the enhanced power focusing effect provided by a larger RIS.

\begin{table}[t]
    \centering
    \caption{Communication performance (MCS index = 7) with different numbers of active RIS rows.}\label{tab:com_rows}
    \vspace{-1ex}
    \begin{tabular}{ c c c c c } %
        \toprule
        Configuration  & 6 rows &  7 rows & 8 rows & Baseline \\ \hline
\hline  
        Ratio (\%) &  74.33 & 78.89 &  82.22 & 59.44  \\ \bottomrule  
    \end{tabular}
    \vspace{-2ex}
\end{table}

\textbf{Impact of antenna's number}.
Increasing the number of antennas typically enhances spatial sensing resolution. To investigate whether a larger antenna array benefits the attacker, we simultaneously increase the number of antennas at both the legitimate sensing Rx and the attacker. In practice, this is achieved by aggregating CSI samples from multiple Wi-Fi NICs and merging them post-capture to emulate a larger antenna array.
Fig.~\ref{fig:acc_anta} shows the sensing accuracy as the antenna count increases from 1 to 12 (i.e., four NICs). The baseline denotes the attacker’s performance without any defense mechanism.
As expected, the performance of both the legitimate sensing Rx and the baseline attacker improves with more antennas, eventually reaching a performance plateau. This effect is attributable to the increased spatial diversity. Notably, the performance gap between the legitimate Rx and the baseline attacker remains small, further validating the efficacy of our design.
In contrast, under the protection of \name, the attacker's accuracy remains largely unaffected by the increased number of antennas. This is because the RIS introduces artificial spatial perturbations specifically aligned with the sensing direction, causing the sensing information to become inherently entangled with the injected perturbations. As previously analyzed in Section~\ref{sec:design}, this coupling renders the two components inseparable at the attacker's side. Consequently, even with a larger antenna array providing more spatial observations, the attacker is still unable to isolate valid sensing features from the perturbed CSI. These results confirm that the proposed \name cannot be circumvented simply by scaling up antenna resources.

\begin{figure}[t]
    \centering
    \subfloat[Sensing]{
        \centering
        \includegraphics[width=0.48\linewidth]{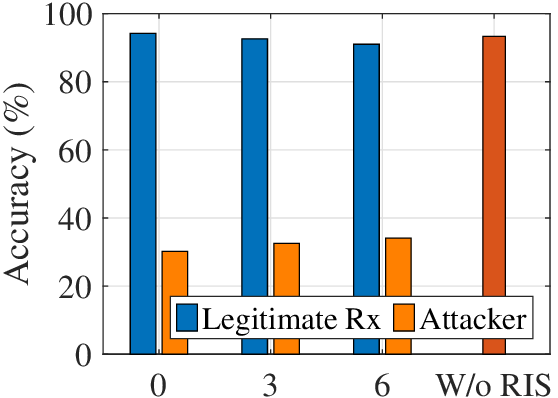}
        \label{fig:acc_error}
    }
    \subfloat[Communication]{
        \centering
        \includegraphics[width=0.48\linewidth]{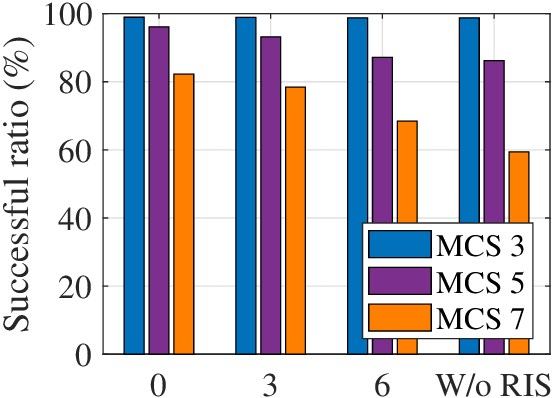}
        \label{fig:com_error}
    }
    \vspace{-1ex}
    \caption{{Impact of angular estimation errors (in degree) on (a) sensing and (b) communication.}}\label{fig:error}
    \vspace{-3ex}
\end{figure}

{\textbf{Impact of angular estimation errors.} To account for potential angular estimation errors in practice, we further evaluate \name under errors of 3$^\circ$ and 6$^\circ$. As shown in Fig.~\ref{fig:error}, \name maintains high communication throughput and sensing accuracy for the legitimate user, while the attacker's performance remains around 30~\!\%. Although a slight degradation is observed as the angular error increases, the overall performance impact is limited. This robustness arises because the RIS-generated beam patterns in Fig.~\ref{fig:beam2} exhibit a relatively wide 3-dB beamwidth (approximately 10$^\circ$), which makes the system tolerant to moderate angular inaccuracies.}

\revhyh{\textbf{Impact of packet rate.} We vary the packet rate from 200~\!Hz to 500~\!Hz to evaluate the sensitivity of demasking, as shown in Tab.~\ref{tab:sen_rate}. The results show that \name maintains high sensing accuracy for the legitimate Rx while keeping the attacker's accuracy low across different packet rates. This is because the packet intervals under these rates are still shorter than the typical indoor channel coherence time, allowing adjacent CSI samples to satisfy the near-static channel assumption needed for reliable demasking.}

\begin{table}[t]
    \centering
    \caption{\revhyh{Sensing performance (\%) with different packet rates.}}\label{tab:sen_rate}
    \vspace{-1ex}
    \begin{tabular}{ >{\centering\arraybackslash}m{4cm} >{\centering\arraybackslash}m{0.6cm} >{\centering\arraybackslash}m{0.6cm} >{\centering\arraybackslash}m{0.6cm} >{\centering\arraybackslash}m{0.6cm} } %
        \toprule
        Packet rate (Hz)  & 200 &  300 & 400 & 500 \\ \hline
\hline  
         Legit.~Rx (W/ \name) &  93.74 & 93.99 &  94.01 & 94.18  \\ 
        Attacker  (W/ \name)  &  29.54 & 30.26 &  30.31 & 30.21 \\ 
        Attacker (W/o \name) &   92.12 &  92.34 &   92.87 & 93.30  \\ 
        \bottomrule  
    \end{tabular}
    \vspace{-2ex}
\end{table}

\subsection{Extended Experiments and Discussion}
\begin{figure}[t]
    \centering
    \vspace{-0.5em}
    \subfloat[Classifier in SignFi]{
        \centering
        \includegraphics[width=0.48\linewidth]{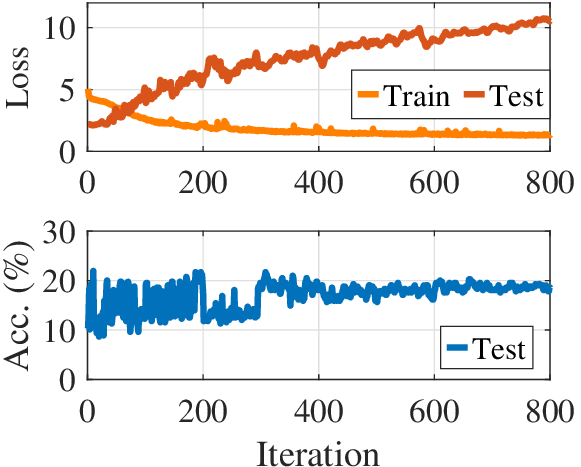}
        \label{fig:loss_cnn}
    }
    \subfloat[ResNet]{
        \centering
        \includegraphics[width=0.48\linewidth]{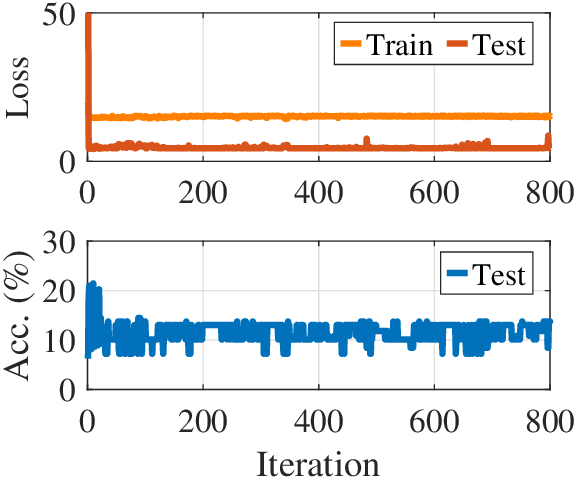}
        \label{fig:loss_resnet}
    }
    \vspace{-1ex}
    \caption{Training loss and testing loss/accuracy under (a) the classifier adopted in SignFi and (b) ResNet.}
    \label{fig:loss_model}
    \vspace{-2ex}
\end{figure}

\textbf{Can the attacker succeed with a self-trained model?}
In previous experiments, both the attacker and the legitimate sensing Rx used the same classifier that is trained on CSI with masking. 
\revhyh{One might argue that the attacker's poor performance could stem from using a model that does not generalize well to the attacker's own raw eavesdropped CSI observations. To evaluate whether an attacker can bypass \name by training an adapted model, we consider an extreme case where the attacker independently collects a training dataset of raw CSI without demasking, e.g., through long-term eavesdropping, and trains a classifier using this dataset.}
As shown in Fig.~\ref{fig:loss_cnn}, when using the SignFi classifier, although the training loss steadily decreases with the SignFi classifier, the test loss continues to increase and the accuracy stays below 20~\!\%, highlighting a clear failure to learn meaningful representations from the CSI.
To rule out the possibility that this poor performance is due to limited model capacity, we further experiment with a more powerful ResNet-based classifier. As depicted in Fig.~\ref{fig:loss_resnet}, even with a deeper model, both training and test losses converge to a non-trivial lower bound, and the test accuracy remains consistently low.
\revhyh{This outcome stems from the fact that, without access to the shared secret key, the attacker observes randomized CSI sequences for the same gesture under the proposed method. Such randomness disrupts the temporal and spatial consistency essential for effective model learning. Moreover, the RIS-induced variations are entangled with human-motion-induced variations and the selected RIS configurations vary over time, making long-term recording ineffective for recovering clean sensing patterns.}
Overall, these findings provide strong evidence of the privacy-preserving effectiveness of our proposed \name, even under enhanced threat models.

\begin{figure}[t]
    \centering
    \subfloat[Respiration waveform]{
        \centering
        \includegraphics[width=0.8\linewidth]{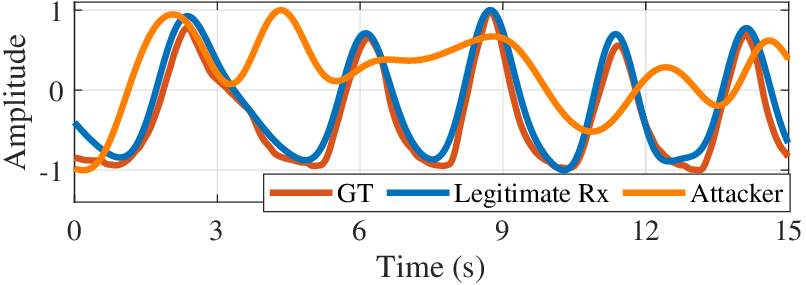}
        \label{fig:fig_waveform}
    }\\
    \vspace{-2ex}
     \subfloat[Legitimate Rx]{
        \centering
        \includegraphics[width=0.48\linewidth]{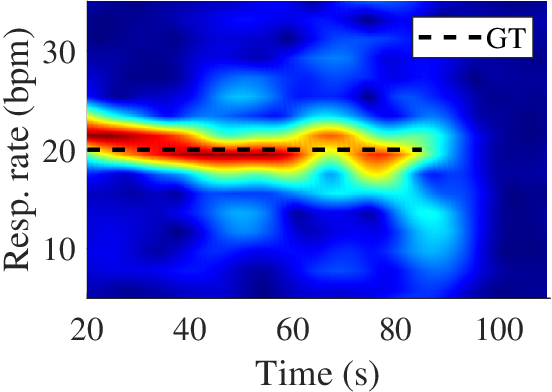}
        \label{fig:fig_bre_rx}
    }
    \subfloat[Attacker]{
        \centering
        \includegraphics[width=0.48\linewidth]{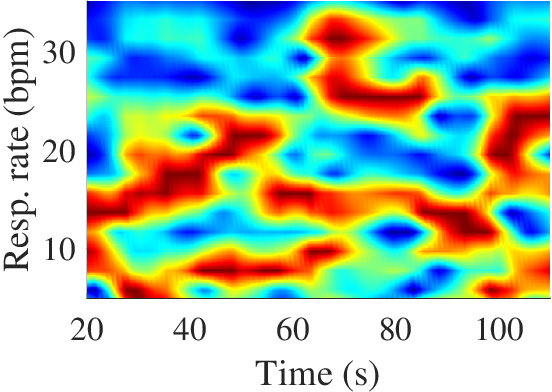}
        \label{fig:fig_bre_attacker}
    }
    \vspace{-1ex}
    \caption{Respiration monitoring: (a) respiration waveform and the spectrograms of (b) the legitimate Rx and (c) the attacker.}
    \label{fig:fig_respiration}
    \vspace{-3ex}
\end{figure}

\textbf{Respiration monitoring.} To further validate the effectiveness of our proposed scheme, we extend the evaluation to a model-based sensing task: respiration monitoring, which relies on signal processing rather than AI-driven classification.
Specifically, we adopt the method from~\cite{zeng2019farsense} to reconstruct the respiration waveform, using a commodity respiratory belt as the ground-truth (GT) reference. As shown in Fig.~\ref{fig:fig_waveform}, the legitimate sensing Rx can accurately recover the waveform. In contrast, the attacker affected by the perturbation introduced by the RIS, fails to reconstruct a reliable waveform.
We further present the corresponding respiration spectrograms in Figs.~\ref{fig:fig_bre_rx} and~\ref{fig:fig_bre_attacker}. During the first 85 seconds, when the volunteer breathes normally, the legitimate Rx successfully tracks the respiration rate, whereas the attacker is unable to do so. In the following 25 seconds, the volunteer holds his/her breath. The legitimate Rx correctly detects the absence of respiration. In contrast, the attacker misinterprets the artificial CSI fluctuations caused by RIS as breathing signals, leading to false detections.
These results demonstrate that \name not only defends against eavesdropping in classification-based applications but also preserves sensing fidelity in model-driven ones. This highlights the broad robustness and practical viability of our design.

\textbf{Discussion.} 
\revhyh{The current leakage level can provide practical protection for password inference, where the attacker must correctly infer multiple consecutive inputs; it can be further reduced by increasing randomized RIS configurations or enlarging the RIS size.}
\revhyh{Secondly, \name can be extended to practical wireless RIS deployments. For example, RISENSE~\cite{deram2025risense} demonstrates low-power wireless RIS control through amplitude-modulated signal decoding, which can be integrated into \name without changing the designed RIS reflection patterns. Moreover, \name does not necessarily rely on directional transmission from the Tx to the RIS. As long as the RIS-induced perturbation is sufficiently strong in the target-reflected sensing path, which can also be achieved by a larger RIS aperture, the proposed design remains effective even with an omnidirectional transmitter.}
\revhyh{Meanwhile, to evaluate \name under a strong attacker, we assume the attacker uses the same sensing model as the legitimate receiver, trained under the same environment; cross-domain sensing generalization is an important issue that we leave for future work.}
\revhyh{Finally, \name can benefit edge-computing and digital-twin systems by protecting wireless sensing measurements before they are used for edge inference, human digital-twin updating, and personalized services~\cite{okegbile2023differentially,yang2024dynamic,chen2024networking,chen2024generative}. This physical-layer protection is also relevant to RIS-assisted digital-twin interaction scenarios~\cite{chen2026generative}, where programmable wireless environments support digital-twin services.}

\revhyh{\textbf{Application scenarios.} \name mainly considers indoor Wi-Fi-based ISAC scenarios, where communication packets from commodity devices are also reused for sensing. Typical examples include smart homes, offices, healthcare and eldercare monitoring, respiration monitoring, gesture or keystroke-related privacy protection, and smart-building sensing applications relying on wireless sensing measurements. In these scenarios, passive eavesdroppers may exploit CSI to infer private target-related information, while legitimate receivers still need reliable communication and sensing. Thus, \name serves as a plug-and-play RIS-assisted privacy layer for such Wi-Fi-based ISAC deployments.}

\section{Conclusion} \label{sec:conclusion}

In this paper, we have proposed \name, a general and practical system that leverages RIS to simultaneously enable high-performance communication and sensing while preserving user privacy. Our design introduces a novel RIS beamforming design that generates two distinct beamforming vectors per RIS row, maximizing signal variation in the sensing direction while maintaining stable, high-gain transmission in the communication direction. By switching between these vectors, \name introduces artificial perturbations that effectively obscure sensitive sensing information.
To enable legitimate sensing, we further develop a time-domain masking and demasking method, allowing only authorized Rx to identify and extract valid sensing information. Experiment results with commodity wireless devices demonstrate that \name provides \revhyh{effective privacy protection} while maintaining high-performance communication and sensing, confirming the effectiveness. With a lightweight implementation, full compatibility with commodity wireless hardware, and ease of deployment, \name serves as a broadly applicable solution for diverse ISAC scenarios.

\bibliographystyle{IEEEtran}
\bibliography{ref}

\end{document}